\documentclass[fleqn,usenatbib,useAMS]{mnras}

\usepackage{graphicx}	
\usepackage{amsmath}	
\usepackage{amssymb}	
\usepackage{multicol}        
\usepackage{bm}		
\usepackage{pdflscape}	
\usepackage{float} 
\usepackage{multirow} 

\newcommand{\sage}{\textsc{SAGE}}
\newcommand{\getemlines}{\textsc{get\_emlines}}

\newcommand{\RawMod}{\textit{RawELGs}}
\newcommand{\DustMod}{\textit{DustELGs}}
\newcommand{\PozModA}{\textit{PozMod1}}
\newcommand{\PozModC}{\textit{PozMod3}}
\newcommand{\PozModAraw}{\textit{RawELGs-Poz1}}
\newcommand{\PozModCraw}{\textit{RawELGs-Poz3}}
\newcommand{\PozModAdust}{\textit{DustELGs-Poz1}}
\newcommand{\PozModCdust}{\textit{DustELGs-Poz3}}

\newcommand{\Halpha}{{\ifmmode{\textrm{H}{\alpha}}\else{H${\alpha}$}\fi}}
\newcommand{\Mbnd}{{\ifmmode{M_{\rm bnd}}\else{$M_{\rm bnd}$}\fi}}
\newcommand{\Mfof}{{\ifmmode{M_{\rm fof}}\else{$M_{\rm fof}$}\fi}}
\newcommand{\Mcrit}{{\ifmmode{M_{\rm 200c}}\else{$M_{\rm 200c}$}\fi}}
\newcommand{\Rcrit}{{\ifmmode{R_{\rm 200c}}\else{$R_{\rm 200c}$}\fi}}
\newcommand{\Rhost}{{\ifmmode{R_{\rm host}}\else{$R_{\rm host}$}\fi}}
\newcommand{\Mmean}{{\ifmmode{M_{\rm 200m}}\else{$M_{\rm 200m}$}\fi}}
\newcommand{\MBN}{{\ifmmode{M_{\rm BN98}}\else{$M_{\rm BN98}$}\fi}}

\newcommand{\hGpc}{{\ifmmode{h^{-1}{\rm Gpc}}\else{$h^{-1}$Gpc}\fi}}
\newcommand{\hMpc}{{\ifmmode{h^{-1}{\rm Mpc}}\else{$h^{-1}$Mpc}\fi}}
\newcommand{\hkpc}{{\ifmmode{h^{-1}{\rm kpc}}\else{$h^{-1}$kpc}\fi}}
\newcommand{\hMsun}{{\ifmmode{h^{-1}{\rm {M_{\odot}}}}\else{$h^{-1}{\rm{M_{\odot}}}$}\fi}}
\newcommand{\Mstar}{{\ifmmode{M_{*}}\else{$M_{*}$}\fi}}
\newcommand{\Mhalo}{{\ifmmode{M_{\rm Halo}}\else{$M_{\rm Halo}$}\fi}}
\newcommand{\Ngal}{{\ifmmode{N_{\rm gal}}\else{$N_{\rm gal}$}\fi}}
\newcommand{\Norph}{{\ifmmode{N_{\rm orphan}}\else{$N_{\rm orphan}$}\fi}}
\newcommand{\Nxorph}{{\ifmmode{N_{\rm non-orphan}}\else{$N_{\rm non-orphan}$}\fi}}
\newcommand{\Zsolar}{{\ifmmode{Z_{\odot}}\else{$Z_{\odot}$}\fi}}
\newcommand{\Msun}{{\ifmmode{{\rm {M_{\odot}}}}\else{${\rm{M_{\odot}}}$}\fi}}
\newcommand{\ltsima}{$\; \buildrel < \over \sim \;$}
\newcommand{\gtsima}{$\; \buildrel > \over \sim \;$}
\newcommand{\lsim}{\lower.5ex\hbox{\ltsima}}
\newcommand{\gsim}{\lower.5ex\hbox{\gtsima}}

\newcommand{\Tab}[1]{Table~\ref{#1}}
\newcommand{\Sec}[1]{Section~\ref{#1}}
\newcommand{\App}[1]{Appendix~\ref{#1}}
\newcommand{\Eq}[1]{Eq.~(\ref{#1})}
\newcommand{\Fig}[1]{Fig.~\ref{#1}}
\newcommand{\beq}{\begin{equation}}
\newcommand{\eeq}{\end{equation}}

\usepackage[T1]{fontenc}
\usepackage{ae,aecompl}

\usepackage{newtxtext,newtxmath}

\title[UNITSIM Galaxies]{UNITSIM-Galaxies: data release and clustering of emission-line galaxies}

\author[Knebe et al.]
{Alexander Knebe,$^{1,2,3}$\thanks{Contact e-mail: \href{mailto:alexander.knebe@uam.es}{alexander.knebe@uam.es}}
Daniel Lopez-Cano,$^{1,9}$
Santiago Avila,$^{1,4}$
Ginevra Favole,$^{5}$
\newauthor
Adam R. H. Stevens,$^{3}$
Violeta Gonzalez-Perez,$^{1,2}$
Guillermo Reyes-Peraza,$^{1,4}$
\newauthor
Gustavo Yepes,$^{1,2}$
Chia-Hsun Chuang,$^{6}$
Francisco-Shu Kitaura$^{7,8}$
\\
$^{1}$Departamento de F\'{\i}sica Te\'{o}rica, M\'{o}dulo 15, Facultad de Ciencias, Universidad Aut\'{o}noma de Madrid, 28049 Madrid, Spain\\
$^{2}$Centro de Investigaci\'{o}n Avanzada en F\'{\i}sica Fundamental (CIAFF), Facultad de Ciencias, Universidad Aut\'{o}noma de Madrid, 28049 Madrid, Spain\\
$^{3}$International Centre for Radio Astronomy Research, University of Western Australia, 35 Stirling Highway, Crawley, Western Australia 6009, Australia\\
$^{4}$ Instituto de  F\'{\i}sica Te\'{o}rica, (UAM/CSIC), Universidad Aut\'omoma de Madrid, Cantoblanco, E-28049 Madrid, Spain\\
$^{5}$ Institute of Physics, Laboratory of Astrophysics, Ecole Polytechnique F\'ed\'erale de Lausanne (EPFL), Observatoire de Sauverny, 1290 Versoix, Switzerland\\
$^{6}$ Kavli Institute for Particle Astrophysics and Cosmology, Stanford University, 452 Lomita Mall, Stanford, CA 94305\\
$^{7}$ Instituto de Astrof\'{i}sica de Canarias (IAC), Calle V\'{i}a Lactea s/n, 38200, La Laguna, Tenerife, Spain\\
$^{8}$ Departamento de Astrof\'{i}sica, Universidad de La Laguna (ULL), E-38206, La Laguna, Tenerife, Spain\\
$^{9}$ Donostia International Physics Center (DIPC), Paseo Manuel de Lardizabal, 4, 20018 Donostia-San Sebastián, Spain
}

\date{Last updated 2015 May 22; in original form 2013 September 5}

\pubyear{2021}

\begin{document}
\label{firstpage}
\pagerange{\pageref{firstpage}--\pageref{lastpage}}
\maketitle

\begin{abstract}
 New surveys such as ESA's Euclid mission are planned to map with unprecedented precision the large-scale structure of the Universe by measuring the 3D positions of tens of millions of galaxies. It is necessary to develop theoretically modelled galaxy catalogues to estimate the expected performance and to optimise the analysis strategy of these surveys. We populate two pairs of (1\hGpc)$^3$ volume dark-matter-only simulations from the UNIT project 
 with galaxies using the \sage\ semi-analytic model of galaxy formation, 
 coupled to the photoionisation model \getemlines\ to estimate their \Halpha\ emission. 
 These catalogues represent a unique suite that includes galaxy formation physics and -- thanks to the fixed-pair technique used -- an effective volume of $\sim (5h^{-1}\rm{Gpc})^3$, which is several times larger than the Euclid survey. We present the performance of these data and create five additional emission-line galaxy (ELG) catalogues by applying a dust attenuation model as well as adjusting the flux threshold as a function of redshift in order to reproduce Euclid-forecast $dN/dz$ values. As a first application, we study the abundance and clustering of those model \Halpha\ ELGs: for scales greater than $\sim 5\hMpc$, we find a scale-independent bias with a value of $b\sim 1$ at redshift $z\sim 0.5$, that can increase nearly linearly to $b\sim 4$ at $z\sim 2$, depending on the ELG catalogue. Model galaxy properties, including their emission-line fluxes (with and without dust extinction) are publicly available.
\end{abstract}




\section{Introduction}
During the last few decades, numerous projects have been aimed at creating large cartographic maps of galaxies, such as 2dFGRS~\citep{2dFGRS}, SDSS~\citep{SDSS,SDSS_BAO}, WiggleZ~\citep{Drinkwater_2010,WiggleZ_final}, BOSS~\citep{Dawson_2012,BOSS_final}, eBOSS~\citep{Dawson_2016,eBOSSfinal} or DES ~\citep{DES,Abbott_2018}. 
They have been carried out with the objective of trying to better understand the large-scale structure of the Universe, to estimate the different parameters that regulate the formation of structures, to determine the expansion history of the Universe, to study how galaxies form, to reconstruct their star formation histories, and to impose constraints upon different models that currently exist for dark energy and for alternative theories of gravity. While advances have certainly been made, all these grand topics remain open areas of investigation, and likely will for years to come.

New surveys such as Euclid \citep{laureijs2011euclid, Amendola_2013}, the Nancy Grace Roman Space Telescope \citep{spergel2013widefield, spergel2015widefield}, the Dark Energy Spectroscopic Instrument  \citep[DESI,][]{collaboration2016desi}, and the 4-metre Multi-Object Spectroscopic Telescope \citep[4MOST,][]{de_Jong_2012} are planned to map with unprecedented precision the large-scale structure of the Universe by measuring the 3D positions of tens of millions of galaxies. These missions are expected to start operating in the coming years, providing the scientific community with wider, deeper, and more accurate data, which may be used to impose stronger constraints upon theoretical models and to provide more accurate estimates for some of the aforementioned parameters relevant in cosmology. Some of these forthcoming missions (e.g., Euclid) will focus on conducting spectroscopic surveys of galaxies using near-infrared grisms in order to determine the positions of galaxies by observing their emission lines such as \Halpha. The wavelength of the observed emission lines will serve to determine the redshifts of the detected objects. Such observations have already been undertaken in the past. There are, for instance, the High-z Emission Line Survey \citep[HiZELS,][]{Geach2008} and the Wide Field Camera 3 Infrared Spectroscopic Parallels survey \citep[WISP,][]{Atek2010}. The WISP survey, for instance, has been used by \citet{Colbert2013} to measure the number density evolution of \Halpha\ emitters; \citet[][]{Sobral2016} employed the HiZELS data (and additional follow-up observations) to quantify the evolution of the \Halpha\ luminosity function. But all previous efforts lack the volumes to be probed by future missions.

Observational campaigns need to be complemented by cosmological simulations: a cornerstone of large-scale structure analysis. Cosmological simulations inform and validate galaxy clustering models. They are also used to test and optimise different estimators and analysis pipelines, to estimate covariance matrices, and to compare with measurements from data. Smaller scales (i.e. below 1 Mpc) are known to contain many more Fourier modes than larger ones and hence constraining power. However, they are heavily affected by the physics of galaxy formation. Since the spatial volumes that the aforementioned surveys seek to study are notoriously large, it is still necessary to rely on dark-matter-only simulations in which galaxies are introduced in post-processing either by halo occupation distribution \citep[HOD, e.g.][as well as the \href{https://sci.esa.int/s/WnvZn6W}{Euclid \textit{Flagship} mock galaxy catalogue}]{Berlind03}, (sub-)halo abundance matching \citep[SHAM, e.g.][]{Vale2004} or semi-analytic models (SAM) \citep[e.g. the MultiDark-Galaxies,\footnote{Galaxy catalogues based upon three distinct SAMs can be downloaded from \href{https://www.cosmosim.org/cms/documentation/projects/galaxies}{CosmoSim}.}][]{Knebe2018b}. While there are efforts to push the limits of `full physics' hydrodynamical simulations to larger and larger volumes \citep[e.g.][]{Lee2020}, it still remains more feasible to match the volumes that missions like Euclid will cover with gravity-only simulations.

The demand for large volumes modelled with sufficiently high resolution is also the reason why, during the last years, alternatives to running such demanding simulations have been explored. For instance, the technique developed by \cite{Angulo_2016} dramatically reduces the variance arising from the sparse sampling of wavemodes in cosmological simulations. The method uses two simulations that are `fixed' and `paired', i.e. the initial Fourier mode amplitudes are fixed to the ensemble average power spectrum and their phases are shifted by $\pi$. This approach has been adopted by the UNIT collaboration\footnote{\url{http://www.unitsims.org}} \citep{Chuang_2019} where it has been shown that the effective volume of such fixed-and-paired simulations can be several times larger than the actual volume simulated: in \citet{Chuang_2019} we have shown that the original four $(1\hGpc)^3$ simulations correspond to a total effective volume of ca. $(5\hGpc)^3$, i.e. $\sim$~7 times of the survey volume of Euclid or DESI. We use the same two pairs of simulations for our study here. Our simulations include the large scales with an accuracy greater than expected by these surveys, and here we have populated them with galaxies using a semi-analytical model that includes all the relevant physical processes for galaxy formation. In terms of galaxy clustering statistics, each pair can be as precise on (non-)linear scales as an average over approximately 150 traditional simulations. They therefore are suitable to statistically study matter--galaxy interplay and galaxy clustering alongside its bias. 

In this work we present and use galaxy catalogues for simulations that were generated by applying the \sage\ semi-analytic model \citep[][]{Croton16} to the aforementioned gravity-only UNIT simulations. These \sage\ galaxies have then been processed with the \getemlines\ code \citep[][]{Orsi_2014} in order to obtain emission-line galaxies (ELGs). Using the resulting ELG catalogues, we study the predicted number density evolution of \Halpha\ emitters and compare it to other theoretical models as well as observational data. We also generate additional ELG catalogues by imposing certain flux threshold and/or even apply a dust attenuation model. All catalogues are used to study the clustering of our \Halpha\ galaxies and their linear bias with respect to the dark matter field, a quantity first studied by \citet{Kaiser84} for Abell clusters and developed in theoretical detail by \citet{Bardeen86}. The bias is a key parameter and a result of not only halo formation but also the varied physics of galaxy formation that can cause the spatial distribution of baryons to differ from that of dark matter. The bias connects the observed statistics to theoretical predictions and has recently been the target of many theoretical studies in light of ELGs \citep[e.g.][]{Geach2012,Cochrane2017,Merson2019,Tutusaus2020}. Our results add to these and may be used to make forecasts for Euclid and related studies for which both the abundance and bias of \Halpha\ ELGs is an input.

There already exist previous works based upon the UNIT simulations and the modelling of ELGs in them \citep{Zhai_2021, Zhai_2019}. However, the important difference to our work is that in those papers only one of the UNIT simulations has been used, as opposed to all four here. Further, Zhai et al. applied a completely different modelling for the ELGs, namely the \textsc{Galacticus} semi-analytic model \citep[][]{benson_galacticus:_2012}, coupled to the \textsc{CLOUDY} photoionisation code \citep{Ferland13} for the calculation of emission line properties. Further, their dust model was tuned as a function of redshift to match observations of the \Halpha\ luminosity function in the redshift range $z\in [0.8,2.3]$. And while Zhai et al. also studied galaxy clustering in the later work, they have not investigated the bias. Our work therefore extends those previous studies and should be viewed as complementary. We further have made our galaxy catalogues publicly available.

The structure of this article is as follows. In \Sec{sec:data} the methods used to generate the ELG catalogues are presented, namely the $N$-body UNIT simulations (\Sec{sec:UNITSIM}), the \sage\ semi-analytic model (\Sec{sec:SAGE}) and the emission-line modelling (\Sec{sec:get_emlines}). Next, in \Sec{sec:galaxies}, we present a series of figures to validate the galaxy catalogues generated by \sage\ by comparing key properties with observational results. Then in \Sec{sec:ELGs} we examine the validity of the modelling for the emission lines of the galaxies. Afterwards, in \Sec{sec:corr}, the results obtained by studying the two-point correlation function and the bias obtained for the ELGs in the Euclid range of redshifts will be presented. Finally, in \Sec{sec:conclusions}, the conclusions derived from this work will be outlined.

\section{The Methods} \label{sec:data}

\subsection{The UNIT Simulations} \label{sec:UNITSIM}

As a basis for this work, four gravity-only simulations that have been developed within the UNIT project have been employed. The names for the two pairs of simulations that we use throughout this work are UNITSIM1 (U1), UNITSIM1-Inverted Phase (U1IP), UNITSIM2 (U2), and UNITSIM2-Inverted Phase (U2IP). The procedure followed for generating these simulations as well as an analysis of the resulting correlation properties is discussed in \cite{Chuang_2019}. For this particular study we have used the two pairs of simulations in which the code \textsc{Gadget} \citep{Springel_2001} has been used to study the behavior of a total of $4096^3$ particles in a volume of 1$h^{-3}$Gpc$^3$ per simulation, thus obtaining a mass resolution of $1.2\times 10^{9}\hMsun$ per simulation particle. 

In \cite{Chuang_2019} it is also explained how the \textsc{Rockstar} halo catalogues and the corresponding \textsc{ConsistentTrees} merger trees have been generated for each of the gravity-only simulations using the publicly available codes from \citet{Behroozi_2012}. All the data corresponding to the UNIT simulations are publicly available at \url{http://www.unitsims.org}. By making the galaxy catalogues and their emission-line properties available too, this work further adds to the community.\\


\subsection{Semi-analytic galaxy modelling via \sage} \label{sec:SAGE}


\sage\ \citep[Semi-Analytic Galaxy Evolution,][]{Croton16} is a modular, publicly available\footnote{\url{https://github.com/darrencroton/sage}} semi-analytic model of galaxy formation, branched from the Munich family of models \citep[specifically from][]{Croton06}. Haloes (in this case, from the UNIT simulations) are initially seeded with ‘hot’ gas based on the cosmic baryon fraction (modulo a reionization factor at higher redshift and in low-mass haloes). Cooling/accretion of this gas onto the central galaxy is based on the two-mode (hot and cold) model of \citet{white91}. Star formation in the disc occurs once the gas is above a critical average surface density \citep[see ][]{Kennicutt89,Kauffmann96}. Metals are immediately injected and gas recycled into the inter-stellar medium (ISM), where a constant mass-loading factor is also applied to reheat gas out of the disc, some of which will end up in an ejected component if the energy budget allows it. A parametrized fraction of the ejected gas (connected to the virial velocity) is reincorporated into the halo on a dynamical time-scale. Satellite galaxies are tracked in the merger trees until merged or unresolved.  Once their subhaloes become unresolved, satellites are either disrupted (where their baryons are placed in intracluster reservoirs) or immediately merged with the central, dependent on how long they survived as a satellite. \sage, therefore, does not have orphan galaxies.  Mergers and disc instabilities trigger starbursts, drive stars into the bulge, and cause gas to be accreted onto the central black hole.  This triggers quasar-mode active galactic nuclei (AGN) feedback, which reheats gas from the disc.  When galaxies have sufficiently (super)massive black holes, cooling is also suppressed by radio-mode AGN activity (both past and present), modelled by a phenomenological ‘heating’ radius that can only grow with time, within which gas cannot cool.

This is the same \sage\ model that was also applied to the MultiDark simulation MDPL2 \citep{Knebe2018b}. The model was calibrated for that simulation by fitting visually first the $z=0$ stellar mass function \citep{Baldry08}, and secondarily using the stellar metallicity--mass relation \citep{Tremonti04}, baryonic Tully--Fisher relation \citep{Stark09}, black hole--bulge mass relation \citep{Scott13}, and cosmic star formation rate density \citep{Somerville01}. The model has not been re-calibrated here as both the UNIT and MDPL2 simulations were run with the same cosmological parameters \citep[][]{Planck2015} and have the same box size. However, the mass resolution is marginally better for UNITSIM, due to the 20 per cent larger number of particles. For the general performance of the \sage\ model we refer the reader to the results presented in \citet{Knebe2018b}, as the calibration plots change minimally when going from MPDL2 to UNIT (see also \Fig{fig:SMF}, in this paper). The calibration does not include constraints for emission-line galaxies.

For a more detailed description of the model we refer the reader to \citet{Croton16} and section~2.4 of \citet{Knebe2018b}.

\subsection{Emission-line galaxy modelling} \label{sec:get_emlines}
Once we have populated the dark matter haloes from the UNIT simulations with the semi-analytic galaxies generated by \sage\ we obtain values for the intensity of the most relevant emission lines such as \Halpha, [OIII]4959, [OIII]5007, [NII]6548 and [NII]6584 for each of the model galaxies. In this study we focus on the \Halpha\ line -- with a particular focus on the Euclid mission. The other emission lines are left for future work.\\

\paragraph*{\getemlines\ code.}
In order to reproduce the intensity of \Halpha\ emission lines of our galaxies, we have used the method presented in~\cite{Orsi_2014}, i.e. the publicly available \getemlines\ code.\footnote{\url{https://github.com/aaorsi/get_emlines}} This code is based on the algorithm MAPPINGS-III described in~\cite{Groves_2004} and~\cite{Allen_2008}, which relates the ionization parameter of gas in galaxies, $q$, to their cold-gas metallicity $Z_{\rm{cold}}$ as:
\begin{equation}
    q(Z)=q_0\left(\frac{Z_{\rm{cold}}}{Z_0}\right)^{-\gamma},
    \label{eq:ionpar}
\end{equation}
where $q_0$ is the ionisation parameter of a galaxy that has cold gas metallicity $Z_0$ and $\gamma$ is the exponent of the power law. We adpoted the suggested values of $q_0=2.8\times10^7\,\rm{cm\,s^{-1}}$ and $\gamma=1.3$, which were found to yield \Halpha\ luminosities for  star-forming galaxies in good agreement with observations \citep{Orsi_2014}. Cold gas metallicity is defined as the ratio between the cold gas mass in metals to the total cold-gas mass:
\begin{equation}
Z_{\rm{cold}}=\frac{M_{Z,\rm{cold}}}{M_{\rm{cold}}}.
\label{eq:metallicity}
\end{equation}

The other relevant component is the star formation rate (SFR).\footnote{Total SFR in \sage\ is the sum of the {\tt SfrDisk} and {\tt SfrBulge} fields.} Note that \sage\ provides this quantity averaged over the previous time-step in the merger trees, despite this interval being broken into sub-time-steps in the code. But the model ideally requires as inputs the instantaneous SFR and cold gas metallicity of galaxies. However, \citet[][]{Favole2020} have shown that for galaxies that are not too bright the differences are negligible. To be able to properly compare our results to observations, we convert the luminosities to fluxes and also apply a dust extinction to the luminosities of the model galaxies. 

Please note that when applying the \getemlines\ code to the \sage\ catalogues,\footnote{We are using the plural here when referring to the catalogues as we will always have at our disposal the four catalogues coming from the two pairs of UNIT simulations.} we rejected all galaxies with a star formation rate equal to zero. One might be inclined to therefore claim that our emission-line galaxies are `star-forming galaxies', but usually a threshold on the specific star formation of order 0.01/Gyr (and hence clearly larger than 0) is assumed to separate `passive' and `star-forming' galaxies. Therefore, our ELGs are based upon \sage\ galaxies that do form stars, but also include `passive' galaxies in the conventional sense.

\paragraph*{Dust extinction.}
We use here a Cardelli extinction law implemented following \citet{Favole2020}\footnote{\url{https://github.com/gfavole/dust}}, but we also summarize it here. The attenuation from interstellar dust is added to the intrinsic \Halpha\ luminosity using:
\begin{equation}
    L(\lambda_j)^{\rm{att}}=L(\lambda_j)^{\rm{intr}}10^{-0.4A_\lambda(\tau_\lambda^z,\theta)},
    \label{eq:attenuation}
\end{equation}
where the attenuation coefficient, as a function of the galaxy optical depth $\tau_\lambda^z$ and the dust scattering angle $\theta$, is defined as \citep[]{Osterbrock1989, Draine2003, Izquierdo-Villalba2019, Favole2020}:
\begin{equation}
A_\lambda(\tau_\lambda^z,\theta)=-2.5\log_{10}\frac{1-\exp(-a_\lambda\,\sec\theta)}{a_\lambda\,\sec\theta}.
\label{eq:attcoeff}
\end{equation} 
In \Eq{eq:attcoeff}, $a_\lambda=\sqrt{1-\omega_\lambda}\tau_\lambda^z$, and $\omega_\lambda$ is the dust albedo. We assume $\cos\theta=0.30$ and $\omega_\lambda=0.56$, meaning that the scattering is not isotropic but forward-oriented, and about 60 per cent of the extinction is caused by scattering.\\

The galaxy optical depth is defined as \citep[][]{Hatton03, DeLucia07}:
\begin{equation}
\tau_{\lambda}^z=\left( \frac{A_{\lambda}}{A_V}\right)_{Z_{\odot}}\left( \frac{Z_{\rm{cold}}}{Z_{\odot}}\right)^{1.6}\left( \frac{\langle N_H\rangle}{2.1\times10^{21}\rm{atoms\,\,cm^{-2}}}\right),
\label{eq:tau}
\end{equation}
in terms of the cold gas metallicity $Z_{\rm{cold}}$ defined in \Eq{eq:metallicity} and the extinction curve for solar metallicity: $Z_{\odot}=0.0134$ \citep[]{Asplund09}.  We assume the \citet[][]{Cardelli1989} extinction law:
\begin{equation}
\left(\frac{A_\lambda}{A_V}\right)=a(x)+b(x)/R_V, 
\label{eq:extcurve}
\end{equation} 
where $x\equiv\lambda^{-1}$, $R_V\equiv A_V/E(B-V)=3.1$ is the ratio of total to selective extinction for the diffuse interstellar medium in the Milky Way, and
\begin{equation} \label{eq:dust_parameters}
\begin{aligned}
a(x)=&1+0.17699\,y-0.50447\,y^2-0.02427\,y^3+\\
      &0.72085\,y^4+0.01979\,y^5-0.77530\,y^6+0.32999\,y^7,\\
      b(x)=&1.41338\,y+2.28305\,y^2+1.07233\,y^3-5.38434\,y^4\\
      &-0.62251\,y^5+5.30260\,y^6-2.09002\,y^7,
\end{aligned}
\end{equation} 
with $y=(x-1.82)$.
The quantity $\langle N_H\rangle$ in \Eq{eq:tau} is 
the mean hydrogen column density defined as \citep[]{Hatton03, DeLucia07}:
\begin{equation}\label{eq:columndens}
\langle N_H\rangle=\frac{M_{\rm{cold}}^{\rm{disc}}}{1.4\,m_{\rm p}\,\pi\,(1.68 R_{1/2}^{\rm{disc}})^2}\,\,{\rm{atoms\,\,cm^{-2}}},
\end{equation}
where $M_{\rm{cold}}^{\rm{disc}}$ is the cold-gas mass of the disc, $m_{\rm p}=1.67\times10^{-27}\,{\rm{kg}}$ is the proton mass, and $R_{1/2}^{\rm{disc}}$ is the half-mass radius of the disc.

We caution that emission lines are expected to be more attenuated than the continuum, e.g. \citet[][]{deBarros2016}, which is the model used here. 

\section{The \sage\ galaxies} \label{sec:galaxies}
The aim of this section is to validate how well our theoretically modelled \sage\ galaxies perform with respect to the quantities that enter into the calculation of the emission-line properties. This involves a) stellar mass, b) star formation rates, c) metallicities, and d) disc lengths. We will further focus on redshifts in the range $z\in [1,2]$ and compare to observational data where possible. For comparisons of other properties to observations and the calibration plots, respectively, we refer the reader to \citet{Knebe2018b} where \sage\ has been applied to the MultiDark simulation MDPL2. Note that in this Section we are using the complete \sage\ galaxy catalogue, not restricting any results to ELGs. However, we provide in the Appendix all the corresponding plots for our model ELGs.

\subsection{Stellar Mass Function} \label{sec:smf}
The stellar mass function (SMF) is one of the most significant properties that can be inferred from galaxy surveys since this function represents the number of galaxies in stellar-mass bins, normalized to the volume of the survey/simulation and to the bin width. Its simplicity yet fundamental importance resides in the fact that the SMF is often employed for calibrating semi-analytic models such as \sage\ used here.

\begin{figure}
\includegraphics[width=\columnwidth]{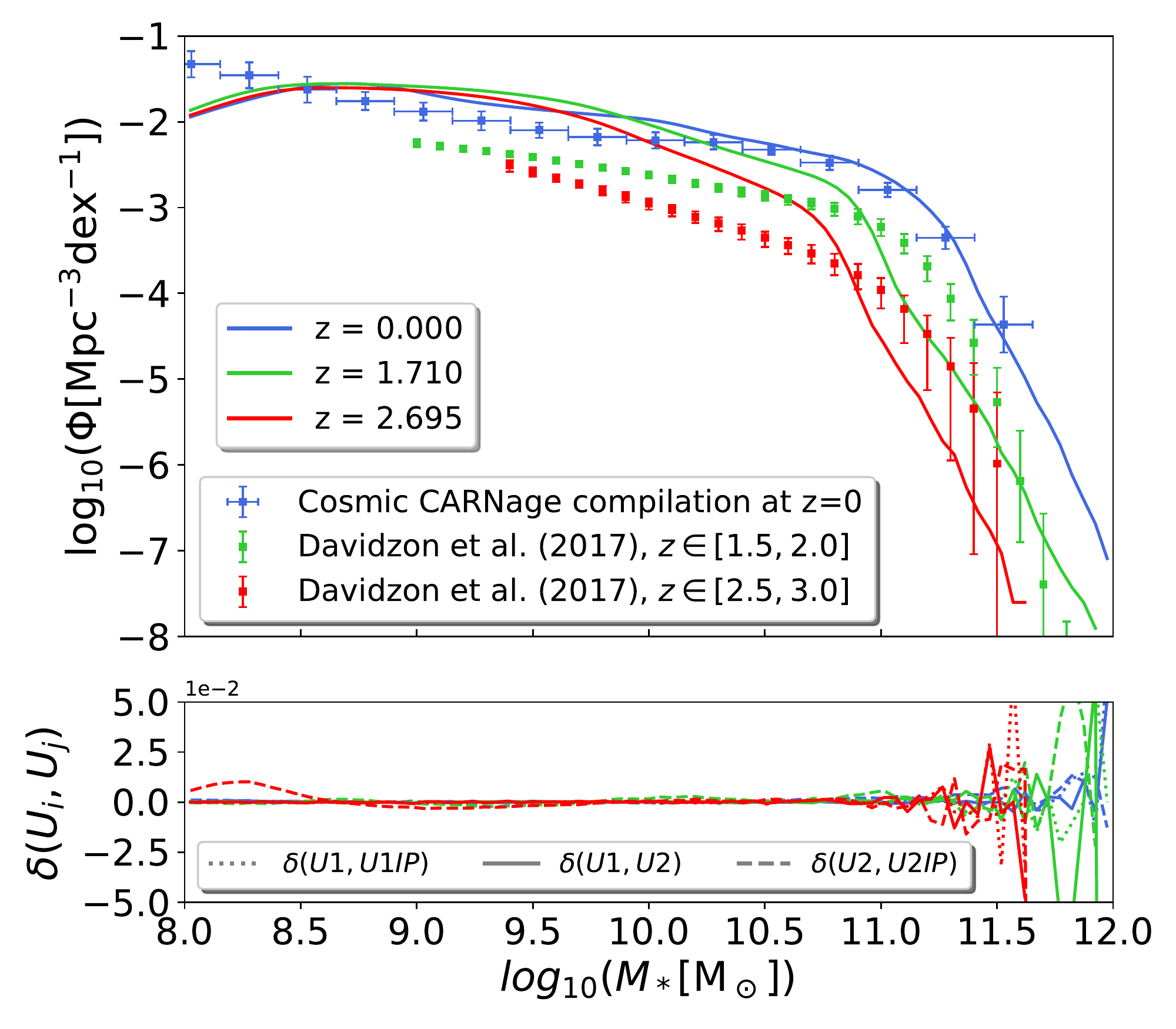}\vspace*{-0.2cm}
\caption{Stellar mass function. In the upper panel we compare the results for the modelled galaxies at various redshifts (solid lines) to observational data (points with error bars). The lower panel shows the fractional difference of U1 to the other UNIT simulations. Note that the $z=0$ SMF has been used to calibrate the \sage\ model whereas the results for higher redshift are a prediction of the model.}
\label{fig:SMF}
\vspace{-0.3cm}
\end{figure}

In the main panel of \Fig {fig:SMF} the results obtained for the SMF computed from the \sage\ galaxies modelled over the UNITSIM1 simulation are presented for three different redshifts $z = [0.0, 1.710, 2.695]$. Together with the results obtained from our simulation, a series of observational results obtained for a range of redshifts similar to those simulated are also represented in the same figure. The compilation for redshift $z=0$ is taken from the so-called `CARNage calibration' data set described in great detail in section~3.3 and appendix~A of \citet{Knebe2018}\footnote{The `CARNage calibration' set is available for download from \url{http://popia.ft.uam.es/public/CARNageSet.zip}.}. The observations for the higher redshifts are taken from \citet{Davidzon2017} and are based on the UltraVISTA near-infrared survey of the COSMOS field. In the bottom panel of \Fig {fig:SMF} the variation in SMF between UNITSIM1 and the three other UNIT simulations is shown, i.e. the $y$-axis represents\footnote{Note that we use the same strategy for presenting the variations across the four UNIT simulations in practically all plots.}
\begin{align}
    \delta(\textrm{U}_i,\textrm{U}_j) = \frac{\textrm{SMF}(\textrm{U}_i)}{\textrm{SMF}(\textrm{U}_j)} - 1,
\end{align}

\noindent
where U$_{i}$ refers to the one of our four UNIT simulations (and U$_{j}$ to another, different one).

For all the simulations conducted, the results produced for the SMF qualitatively follow the observational trends. This outcome is in line with previous results such as those presented in \citet{Favole2020} and \citet{asquith2018cosmic}. The results obtained at redshift $z=0$ agree almost seamlessly with the observational data. This is readily explained by the fact that the \sage\ model was pre-calibrated to very similar data. When studying the behavior at higher redshifts (which is a prediction of the model) certain discrepancies start to show up. For stellar masses below $10^{11} \Msun$ the SMF calculated for the \sage\ galaxies exceeds the observational points, while the opposite is true for masses higher than $10^{11} \Msun$. This is related to the condition that getting both the SMF at $z=0$ \textit{and} the cosmic star-formation history to simultaneously agree with the observations demands that stars that should have been formed in haloes below this simulation's resolution limit must instead be formed as extra stars in the haloes that are resolved. This inevitably leads to resolved high-$z$ galaxies having too much stellar mass (and star-formation rates that are too high) in the model. It also changes how galaxies acquire stellar mass through mergers (as fewer mergers are resolved), which might help explain why there are too few galaxies with $M_*>10^{11}$\hMsun\ at higher $z$ in the model. Additionally, the deviations observed here for high redshifts -- especially at the low-mass end -- are similarly found when studying the SMF produced by other semi-analytic models, as extensively discussed in~\citet{asquith2018cosmic}. Our explanation is hence generic and not only applies to \sage. Therefore, despite the discrepancies seen in \Fig{fig:SMF}, the results obtained are reasonably accurate for us to say that the modelled \sage\ galaxies fairly depict the behaviour of the SMF that could be expected in the redshift range for which Euclid is planned to operate.

Another important aspect worth mentioning in this section is that due to resolution limitations in our simulations, galaxies whose stellar mass is lower than $10^9 \Msun$ have not been considered. Please refer to \citet{Knebe2018, Knebe2018b} for a justification of this threshold, but we can also see in \Fig{fig:SMF} how the number of galaxies starts to decline for stellar masses below that threshold due to numerical limitations. Therefore, to produce the results presented in the following sections we will discard all those galaxies whose mass is inferior to this threshold. This is not a cause for concern in this work though, as the vast majority of relevant ELGs have stellar masses above this threshold (see \App{app:SMFofELGs}).

\subsection{Star Formation} \label{sec:starformation}
\begin{figure}
\includegraphics[width=\columnwidth]{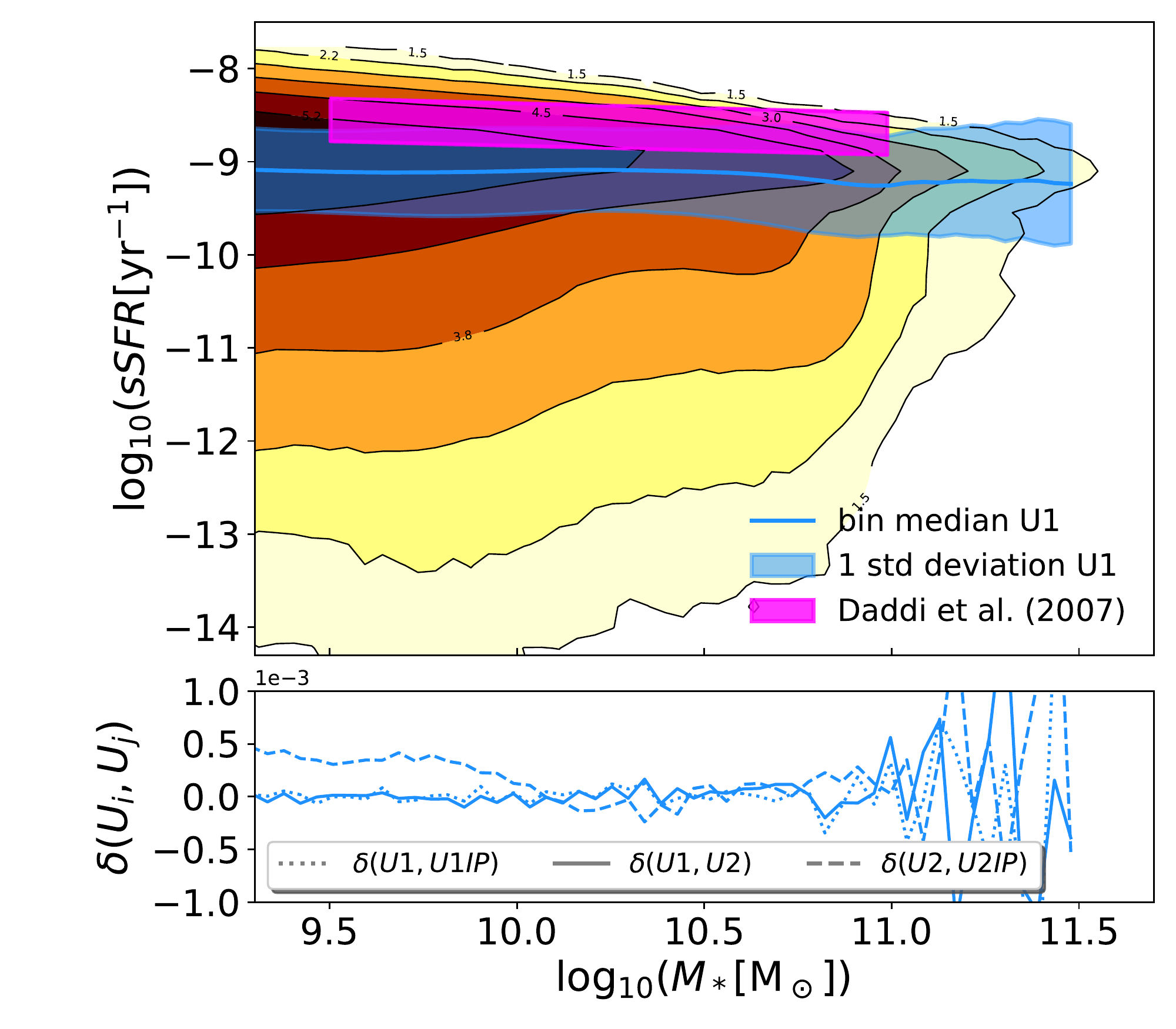}
\caption{Specific star formation rate vs. stellar mass at redshift $z\sim 2$ in comparison to observations by Daddi et al. (2007). 
We show both the median (solid blue line) and the contour levels of the scatter relation. This relation is a prediction of the \sage\ model.}
\label{fig:sSFR2028}
\end{figure}

With respect to the star formation (SF) in galaxies, which is also used as an input to the \getemlines\ code, we only present the relation between specific star formation (i.e. SF per unit stellar mass) and stellar mass at redshift $z\sim2$. We find that \sage\ makes a prediction for this relation that is in excellent agreement with the observations of \citet{Daddi2007}: in the main panel of \Fig{fig:sSFR2028} the specific SF rate (sSFR) of U1-\sage\ galaxies is plotted against the stellar mass $M_{*}$ for redshift $z=2.028$. We show both the contours of a 2D histogram of this scatter plot as well as the median of the values obtained for sSFR within a series of bins along the $x$-axis. As is customary, in the bottom panel of the \Fig{fig:sSFR2028} the variations between simulations with respect to the other UNITSIM-\sage\ galaxies have been represented. When comparing our results to observational data extracted from \cite{Daddi2007}, we find sufficient agreement, at least within the $1\sigma$ regions. Though not explicitly shown here, we also confirm that our \sage\ results are in excellent agreement with observational data for the sSFR \citep[as provided by][]{elbaz2011goods} as a function of stellar mass at redshift $z=0$. These results, in turn, are also compatible to those shown in~\cite{Favole2020} for redshift $z = 0.1$.

For a comparison of the star formation rate (SFR) function to observational data at redshift $z=0.14$ and the redshift evolution of the cosmic star formation rate density, we refer the reader to \citet{Knebe2018b}. While the SFR function is compatible with the observational data at low redshift -- as seen for the MultiDark galaxies and also confirmed for the UNITSIM galaxies (though not explicitly presented here) -- it is worth mentioning that for SFR values greater than $\sim 10^{1.6} \Msun / {\rm yr}$, the number of galaxies generated with \sage\ seems to underestimate the observed number \citep[see fig.~2 in][]{Knebe2018b}. As we will see later in \Sec{sec:abundanceevolution} this is going to leave an imprint on the abundance of (dust-attenuated) ELGs, especially at high redshifts. We finally like to remark again that the relation between sSFR and stellar mass as shown here is a prediction of the \sage\ model.

Based on these results, we can say that our galaxies sufficiently reproduce the behaviour of the sSFR that would be expected for a sample of real galaxies in Euclid's operating range of redshifts.

\subsection{The mass--metallicity relation} \label{sec:zgas2mstar}

Another aspect of galaxies to be considered for the emission-line modelling is the chemical composition, since -- depending on the fraction of metals that a galaxy may contain -- its SFR may be substantially modified due to the fact that a higher metal content favours cooling mechanisms. This property is explicitly taken into account by the \getemlines\ code and has to be provided as an input, respectively.

\begin{figure}
\includegraphics[width=\columnwidth]{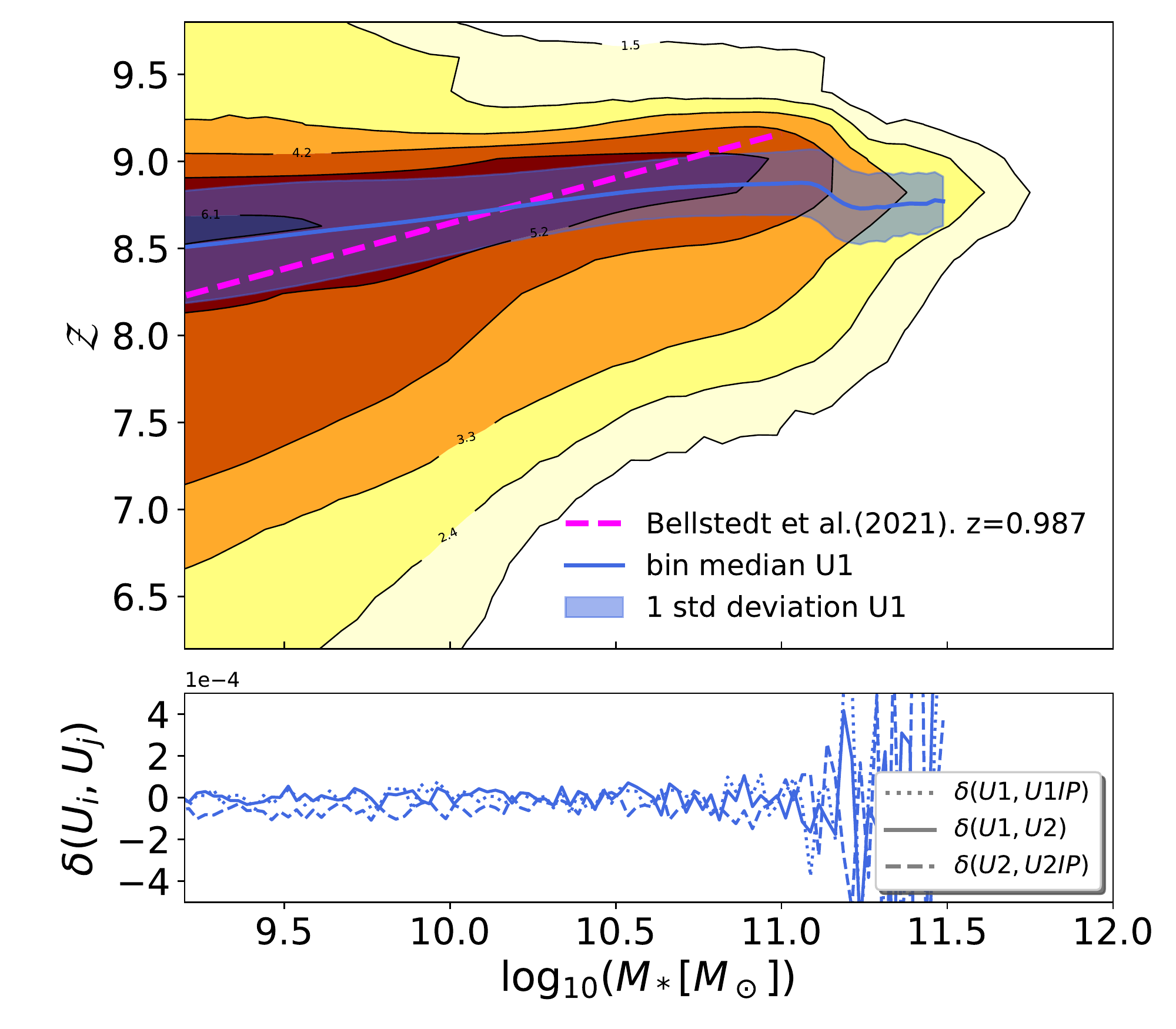}\vspace*{-0.2cm}
\caption{Cold gas metallicity vs. stellar mass. The relation shown here for redshift $z\sim1$ is a prediction of the \sage\ model. We also show the relation as found in \citet{Bellstedt2021} (dashed line).}
\label{fig:Zcold}
\vspace{-0.3cm}
\end{figure}

Since SF is regulated by the collapse of cold gas clouds, in \Fig{fig:Zcold} we study the relation that exists between the total mass of metals contained in such clouds and the total mass of cold gas in a given galaxy throughout the parameter $\mathcal{Z}$ which is calculated as \citep{Favole2020,Knebe2018b}:
\begin{align} \label{eq:Zcold}
    \mathcal{Z} = 8.69 + \log_{10}(Z_{\rm cold}) - \log_{10}(Z_{\odot})  
\end{align}

\noindent
where $Z_{\rm cold}$ was previously defined in \Eq{eq:metallicity}, and we use the same $Z_{\odot}=0.0134$ as already in \Eq{eq:tau}. Note that this quantity $\mathcal{Z}$ is meant to be a proxy for $12+\log ({\rm O/H})$.

In the main panel of \Fig{fig:Zcold} we show the correlation between $\mathcal{Z}$ and stellar mass as contours alongside the median (solid blue line) for redshift $z\sim 1$. The lower panel shows again the fractional difference with respect to to the other UNITSIM model. The relation is as expected, i.e. larger mass galaxies have larger metallicities, with a strength comparable to the one observed for lower redshifts. This relation -- as observed at redhift $\sim0.1$ -- is used during the calibration of the \sage\ model; its extension to $z\sim1$ shown here nevertheless is a clear prediction. We also show the relation as expected from observations by using the best-fit function presented in \citet[][eq.~6]{Bellstedt2021}. This fitting function was obtained by applying the spectral-energy-distribution-fitting code \texttt{ProSpect} to galaxies from the Galaxy and Mass Assembly (GAMA) survey at $z<0.06$; comparing with observations of gas-phase metallicity over a large range of redshifts, they then showed that their best-fit evolving mass--metallicity relationship  is consistent with observations at all epochs and hence used here by us at redshift $z\sim 1$. We only show the Bellstedt et al. function out to $M_*=10^{11}\Msun$ which was their limit for obtaining the best-fit parameters. The predictions of the \sage\ model are in fair agreement with the Bellstedt et al. function. If one were to extrapolate the Bellstedt results, we would find a deficit of cold gas metallicity for the highest mass galaxies with $M_*>10^{11}\Msun$. Even though there is no observational data in that regime, one possible explanation could be that the cold gas in those galaxies comes from mergers rather than accretion/cooling. I.e. AGN feedback might have shut off cooling entirely, so enriched gas in the circumgalactic medium will not get back to the inter-stellar medium. Instead, we might just be seeing the low-metallicity gas from now-cannibalised low-mass galaxies dominating most of the cold gas in the galaxy. But it yet remains unclear if the the drop in metallicity \textit{predicted} for \sage\ galaxies at $M_*=10^{11}\Msun$ will also be seen in observations. While the redshift $z\sim 1$ is relevant for the Euclid mission, it also appears to be important to verify the mass-metallicity relation for even higher redshifts as it plays an important role in the estimation of emission lines. The \citet{Bellstedt2021} function can also be used to obtain results at, for instance, $z=2$. There also exists a best-fit relation derived from actual observations at $z=2.2$ \citep[][eq.~2 together with table~5]{Maiolino2008}. We refrain from showing the corresponding plot here, but confirm that our \sage\ galaxies reproduce those two observations equally well as seen here for $z=1$.

\subsection{The disc size--mass relation} \label{sec:rdisk2mass}

\begin{figure}
\includegraphics[width=\columnwidth]{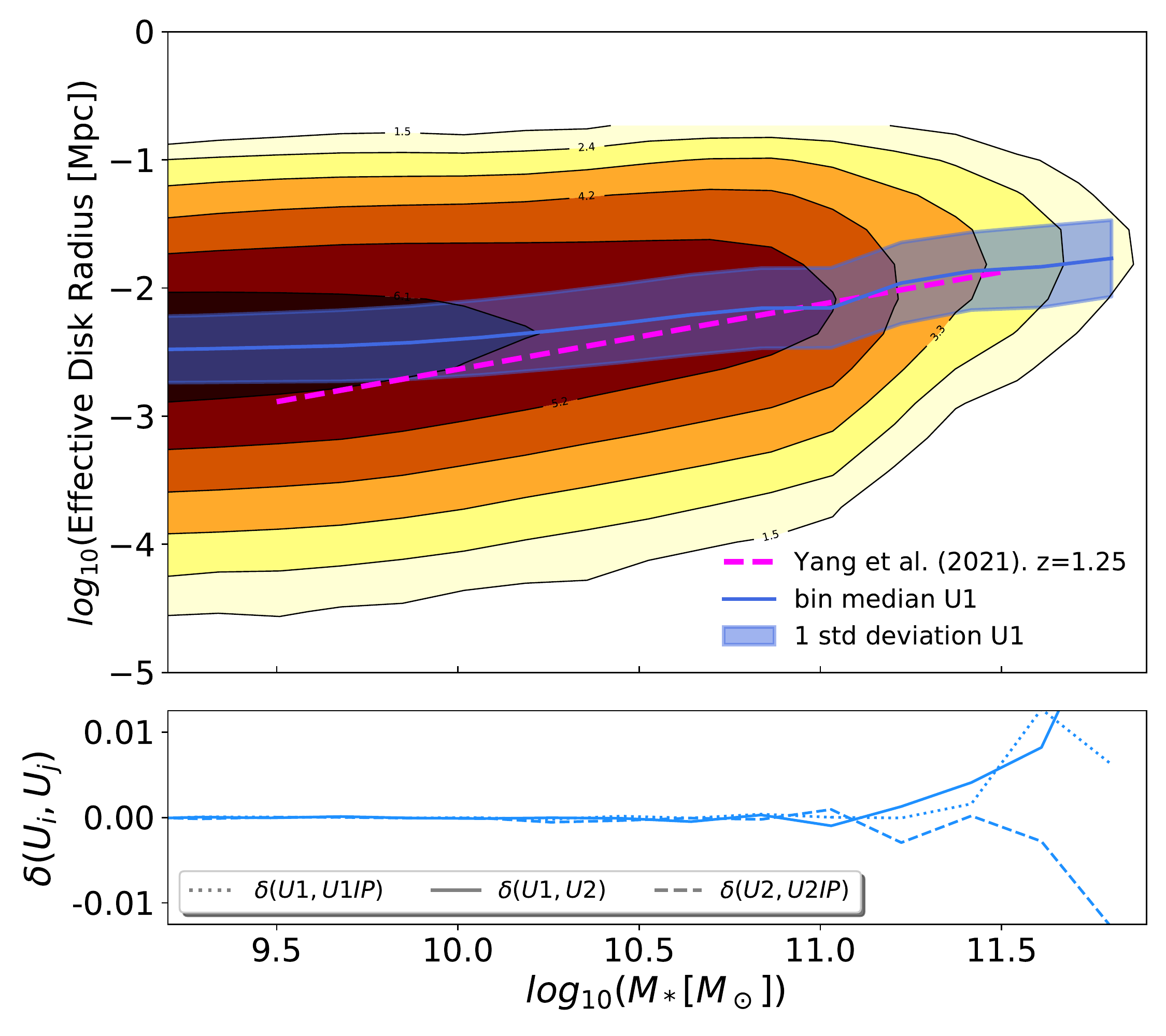}\vspace*{-0.2cm}
\caption{Effective disc radius as a function of stellar mass (contours and blue solid line with 1$\sigma$ error region). This is a prediction of the \sage\ model. We also show the relation as reported for late-type galaxies in \citet{Yang2021} at $z=1.25$ (dashed line).}
\label{fig:R_vs_SM}
\vspace{-0.3cm}
\end{figure}

The last relevant quantity to validate for our \sage\ galaxies is the size of the disc. While it is not important for \getemlines\, it nevertheless enters into our dust attenuation model via \Eq{eq:columndens}. We therefore show in \Fig{fig:R_vs_SM} the correlation of the effective disc radius (i.e. exponential scale radius, as calculated by \sage) with stellar mass at redshift $z=1.22$. For comparison we use the best-fit relation as reported by \citet[][eq.~1]{Yang2021} for late-type galaxies at redshift $z=1.25$ and as derived from the complete Hubble Frontier Fields data set. While the agreement is very good for higher mass galaxies, the disc sizes predicted by \sage\ for galaxies with mass $M_*<10^{10}\Msun$ are systematically larger than the observed ones. However, this does not significantly affect our results here as our ELGs preferentially have stellar masses $M_*>10^{10.5}\Msun$ (see \Fig{fig:SMFofELGs}).\\

Given all the results presented throughout this particular section, with the majority even being predictions of the \sage\ model, we are confident that our UNITSIM-\sage\ galaxies meet all the requirements to be used for the emission-line modelling, which is discussed in great detail in the following section. 

\section{\sage's Emission-Line Galaxies (ELGs)}\label{sec:ELGs}
The results presented in the previous section indicate that our \sage\ model galaxies are in sufficient agreement with a range of observations, in particular those properties that are used as an input for the model that calculates spectral emission lines. Here we now focus on the ELGs and contrast additional properties with a set of observations.\footnote{The same validation plots as shown in \Sec{sec:galaxies} for the \sage\ galaxies can be found for the ELGs in \App{app:ELGs}.} To this extent, we start with generating two distinct ELG catalogues, constructed from the full list of \sage\ galaxies: one set will be obtained by simply applying the \getemlines\ code (\RawMod) and another one by additionally modelling dust extinction (\DustMod). These value-added properties are included in the publicly available catalogues. However, in order to compare to existing observations and to make predictions for Euclid, we apply a redshift-independent flux cut of $F_{\rm cut}=2\times 10^{-16}$~erg s$^{-1}$ cm$^{-2}$, which corresponds to the limit of the Euclid satellite. 

 \begin{figure}
   \includegraphics[width=\columnwidth]{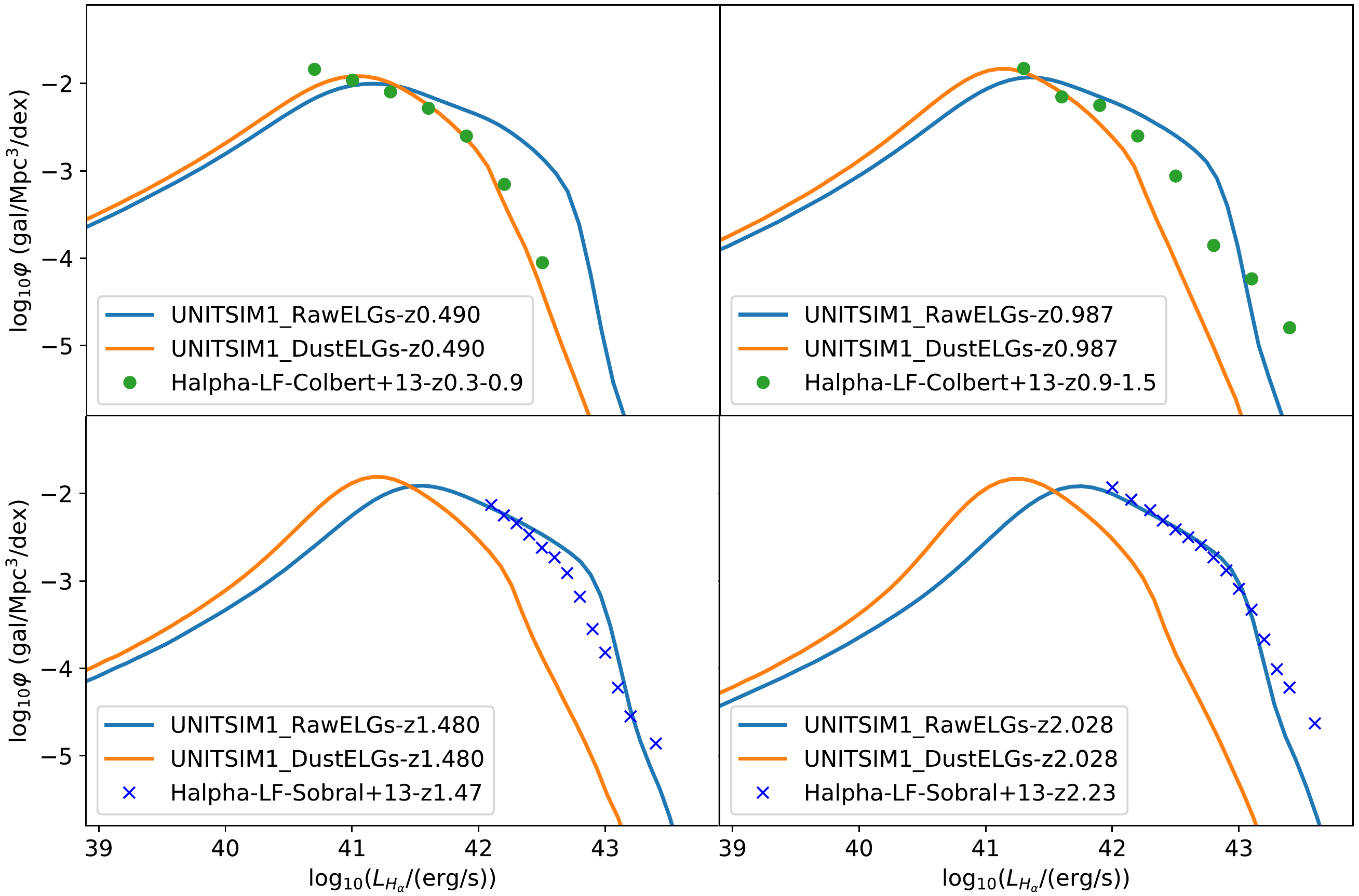}
   \caption{Evolution of the \Halpha\ luminosity function for our \RawMod\ (blue) and \DustMod\ catalogues (orange) ELGs in comparison to observational data: the \citet{Colbert2013} data (red points) are best compared to \DustMod\ whereas the \citet{Sobral2016} data (blue crosses) to \RawMod\ (see main text for details). Redshifts are (clockwise starting in the upper left panel) $z=0.490, 0.987, 1.480,$ and $2.018$.}
 \label{fig:LF}
 \end{figure}

\subsection{The luminosity function of \Halpha-ELGs}\label{sec:LFOfELGs}
 We start with comparing the \Halpha\ luminosity functions (LFs) -- as obtained by \getemlines\ -- at various redshifts of interest to observational data. The results can be viewed in \Fig{fig:LF} for the two base catalogues \RawMod\ and \DustMod\ at $z=0.49, 0.987, 1.48$, and $2.23$. For the first two redshifts we contrast our theoretical LFs to observations as found in \citet{Colbert2013}. The data are taken from their table~3, where we removed again the [NII] contamination; as the data have not been corrected for dust extinction, they are best compared against our \DustMod. For the latter two redshifts, the observations from \citet{Sobral2016} are used. We used the data as provided in their table~4, noting that here they corrected for dust extinction, and hence those curves should be compared against our \RawMod. We actually find that our ELGs match the observations fairly well, though there are some obvious discrepancies at redshift $z\sim 1$: both \DustMod\ and even \RawMod\ do not provide enough high-luminosity ELGs. This eventually translates into a too-low (integrated) abundance, as we will see below. But we are not too concerned about that as the relevant redshift range for Euclid is $z\in [0.9,1.8]$, and the match of our \RawMod\ galaxies with the \citet{Sobral2016} observations is rather good for $z\sim 1.5$, i.e. the centre of that interval. 

In the Introduction we mentioned that \citet{Zhai_2019} also model \Halpha\ ELGs using the \textsc{Galacticus} SAM coupled to the single UNITSIM1 simulation. But their catalogue was constructed such that the SAM parameters were tuned to best reproduce -- amongst other properties -- the \Halpha\ LFs, and in particular the observed ones shown here for redshifts $z=1.48$ and $2.23$ (see their fig.~1). They accomplish this by -- in practice -- adjusting $A_\lambda(\tau_\lambda^z,\theta)$ (as also found in our \Eq{eq:attenuation}) as a free parameter, tuning it until they match the observed \Halpha\ LF at a given redshift. Our value for $A_\lambda(\tau_\lambda^z,\theta)$ is based upon physical properties of the underlying galaxies whose values change as a function of redshift (leading to an implicit redshift dependence of our dust model). Meaning, we actually use a physically motivated $A_\lambda$ and hence the LFs seen here are a clear prediction of our modelling.

\begin{table*}
\caption{Average number density and flux cuts as a function of redshift $z$ (first column). Columns 2--3 (\textit{RawELGs}) and 9--10 (\textit{DustELGs}) list the mean and standard deviation (across the four UNIT simulations) of the number density of ELGs with an applied redshift-independent flux cut of $F_{\rm cut}=2\times 10^{-16}$~erg s$^{-1}$ cm$^{-2}$. Columns 4--5 and 6--7 give the target number density (taken from table~3 in P16) and average flux cut applied to reach it (the standard deviation is smaller than the reported accuracy and hence left out for clarity) for \PozModAraw\ and \PozModCraw, respectively. Columns 10,11 and 12,13 provide the same information for \PozModAdust\ and \PozModCdust.}
\begin{center}
\begin{tabular}{c|cc|cc|cc|c|cc|cc|cc}
\hline
    $z$
    & \multicolumn{2}{c}{\RawMod\ ($F_{\rm cut}=2$)}
    & \multicolumn{2}{c}{\PozModAraw}
    & \multicolumn{2}{c}{\PozModCraw}
    &
    & \multicolumn{2}{c}{\DustMod\ ($F_{\rm cut}=2$)}
    & \multicolumn{2}{c}{\PozModAdust}
    & \multicolumn{2}{c}{\PozModCdust}\\
    \hline
    
    & <$dN/dz$> & $\sigma$ 
    & $dN/dz$   & <$F_{\rm cut}$>
    & $dN/dz$   & <$F_{\rm cut}$> 
    &
    & <$dN/dz$> & $\sigma$
    & $dN/dz$   & <$F_{\rm cut}$>
    & $dN/dz$   & <$F_{\rm cut}$>  \\
\hline
\hline
0.490 & 24652 & 47 & 9946 & 6.441 & --   & --    & & 15262 & 30 &  9946 & 2.85 &  --   & -- \\
0.987 & 22015 & 89 & 7353 & 4.864 & 3779 & 7.080 & &  3238 & 12 &  7353 & 1.37 &  3779 & 1.88 \\
1.220 & 17709 & 94 & 5097 & 4.600 & 2518 & 6.300 & &   957 &  7 &  5097 & 1.13 &  2518 & 1.50 \\
1.321 & 15809 & 98 & 4281 & 4.452 & 2148 & 5.759 & &   577 &  4 &  4281 & 1.05 &  2148 & 1.35 \\
1.425 & 13988 & 77 & 3447 & 4.343 & 1817 & 5.353 & &   370 &  3 &  3447 & 0.98 &  1817 & 1.22 \\
1.650 & 10277 & 57 & 2253 & 3.930 & 1279 & 4.564 & &   153 &  3 &  2253 & 0.84 &  1279 & 1.00 \\
2.028 &  5294 & 38 & 1006 & 3.330 &  616 & 3.687 & &    35 &  1 &  1006 & 0.67 &   616 & 0.77 \\
\hline
\end{tabular}
\label{tab:FluxThresholds}
\end{center}
\end{table*}

\subsection{Abundance evolution of flux-selected \Halpha-ELGs} \label{sec:abundanceevolution}
\begin{figure}
   \includegraphics[width=\columnwidth]{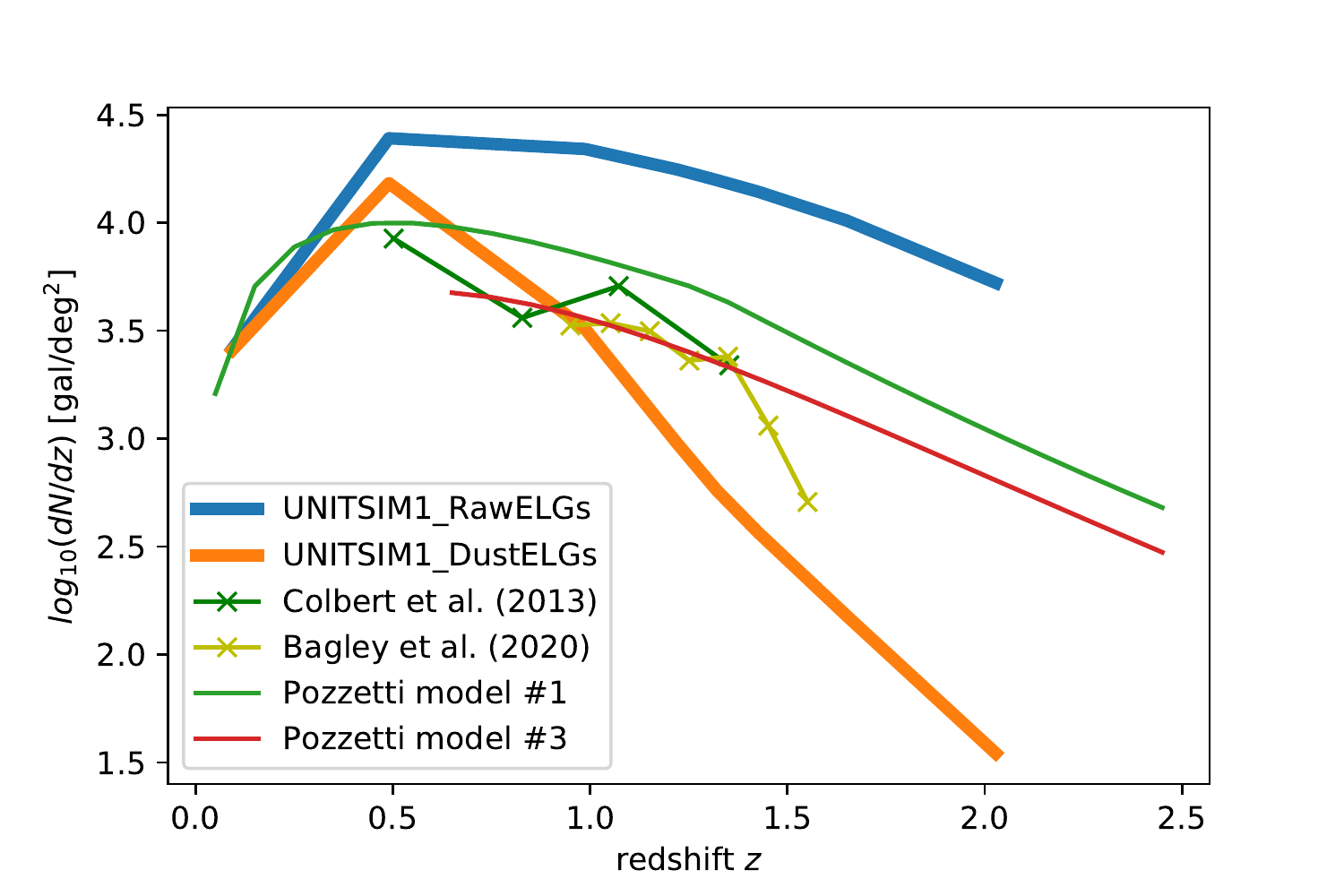}\vspace*{-0.2cm}
   \caption{Redshift evolution of the number density of dust-attenuated ELGs (\DustMod, orange) and the initial \RawMod\ catalogue (i.e. no dust modelling, blue), both for a redshift-independent flux cut at $2\times 10^{-16}$~erg s$^{-1}$ cm$^{-2}$, in comparison to the observational data of \citet[][]{Colbert2013} and \citet{Bagley2020}. We also show model~\#1 and~\#3 of \citet{Pozzetti2016} for the same flux threshold (not to be confused with our catalogues \PozModA\ and \PozModC\ that were designed to match these number densities). Only UNITSIM1 ELGs are shown for clarity.}
 \label{fig:dNdz}
 \vspace{-0.3cm}
 \end{figure}

We show in \Fig{fig:dNdz} the redshift evolution of the number density for our \RawMod\ and \DustMod\ catalogues, after applying the redshift-independent flux cut of $F_{\rm cut}=2\times 10^{-16}$~erg s$^{-1}$ cm$^{-2}$, in comparison to observational data from \citet{Colbert2013} and \citet{Bagley2020}. We also show two of the three models of \citet[][P16]{Pozzetti2016}. By fitting to observed luminosity functions from existing \Halpha\ surveys, P16 build three distinct models for the \Halpha\ number density evolution. Different fitting methodologies, functional forms for the luminosity function, subsets of the empirical input data, and treatment of systematic errors were considered to explore the robustness of the results. Functional forms and model parameters were made available\footnote{The P16 data can be downloaded from here: \url{http://www.bo.astro.it/~pozzetti/Halpha/Halpha.html}} (and are being used here), along with the counts and redshift distributions up to $z\sim 2.5$ for a range of limiting fluxes bracketing the sensitivity of Euclid. Their models are named `Pozzetti model \#1, \#2, and \#3', with model \#1 being the most optimistic and model \#3 the most pessimistic for Euclid.\footnote{P16 called the models that way themselves, based upon the fact that if you have more galaxies, you reduce the shot-noise. Hence, Pozzetti model \#1 is more optimistic for Euclid's figure-of-merits than \#3, as we will have smaller error bars in the cosmological parameters.} Both these models are shown here, also for a flux cut of $2\times 10^{-16}$~erg s$^{-1}$ cm$^{-2}$. 

We can see in \Fig{fig:dNdz} how, for $z<1$, our \DustMod\ follow the same trends as the P16 models, but show a substantial lack of objects at higher redshift. By comparison, our \RawMod\ clearly overpredict the abundance of ELGs for the applied redshift-independent flux cut (at least for $z>0.5$). A similar discrepancy between semi-analytic galaxies and the P16 models can also be seen in fig.~5 of P16, where their three models are compared against the results from two other SAMs. It should also be mentioned that a more recent study of the observed number density evolution of \Halpha\ ELGs indicates a possible decline beyond redshift $z\sim 1.4$ \citep[][lower right panel of their fig.~7]{Bagley2020}, although it is not as pronounced as the dip found for our \DustMod. To highlight this we have added those data points\footnote{The \citet{Bagley2020} data are based upon completeness-corrected measurements of the blended \Halpha\ and NII fluxes, while our fluxes include only \Halpha. We have therefore `corrected' the \citet{Bagley2020} data points -- as obtained with PlotDigitizer -- by reversing their adjustment to model~\#3 of P16 to account for the combined fluxes. This was done by finding the shift needed to bring the digitized data points into the same kind of agreement with the original Pozzetti model \#3, as seen in the lower right panel of Bagley's fig.~7 for the blended Pozzetti model \#3.} to our plot, too. While there is agreement between the observations of Bagley et al. and P16's model~\#3 in the redshift range $z\in [1,1.5]$, the observational data drop more steeply at higher redshifts and are more in line with our \DustMod\ prediction. However, \citet{Bagley2020} also say that their higher redshift points are in the region were the sensitivity of their instrument could be degraded.

This discrepancy between ours and the Pozzetti ELG number densities is also reflected in \Tab{tab:FluxThresholds}, where we list as a function of redshift the number density of ELGs in our reference \RawMod\ and \DustMod\ catalogues (as averaged over the four UNIT simulations, also providing the standard deviation). While we have to acknowledge that both our \RawMod\ and \DustMod\ do not reproduce the predictions of P16, we also have to remark again that it yet remains unclear what the correct abundance evolution $dN/dz$ should look like.

\subsection{Flux-adjusted catalogues} \label{sec:fluxcatalogues}
Taking the models of P16 as the reference, we now construct four additional catalogues that are designed to match the P16 $dN/dz$ curves as shown in \Fig{fig:dNdz}. We take \RawMod\ as the starting point and adjust the flux threshold until reaching the target $dN/dz$ values as given by P16's models \#1 and \#3, providing us with the two models \PozModAraw\ and \PozModCraw. We use the same approach for \DustMod, providing two more models: \PozModAdust\ and \PozModCdust. We used this methodology with all four UNITSIM catalogues. The means of the required flux cuts to our data are listed in columns 4--7, and 10--13 of \Tab{tab:FluxThresholds} (we omit error estimates as they are below the reported accuracy). The remaining columns -- 2, 3, 8, and 9 -- are the mean number densities (and its standard deviation) for the \RawMod\ and \DustMod\ catalogues, respectively, when using a redshift independent flux threshold of $F_{\rm cut}=2\times 10^{-16}$~erg s$^{-1}$ cm$^{-2}$. Our methodology for constructing ELGs eventually leaves us with six distinct catalogues\footnote{We need to state here again that the public versions of \RawMod\ and \DustMod\ are \textit{not} subjected to any flux cut: they contain all ELGs as provided by \getemlines.}

\begin{enumerate}
    \item[\it \textbf{RawELGs}:]  directly coming from \getemlines\ (with a flux threshold of $F_{\rm cut}=2\times 10^{-16}$~erg s$^{-1}$ cm$^{-2}$ across all redshifts, when used here),
    \item[\it \textbf{RawELGs-Poz1}:] variable flux threshold applied to \RawMod\ to match the number density of Pozzetti's model \#1 at each redshift,
    \item[\it \textbf{RawELGs-Poz3}:] variable flux threshold applied to \RawMod\ to match the number density of Pozzetti's model \#3 at each redshift,
    \item[\it \textbf{DustELGs}:] passing the \RawMod\ ELGs through our dust model (with a flux threshold of $F_{\rm cut}=2\times 10^{-16}$~erg s$^{-1}$ cm$^{-2}$ across all redshifts, when used here),
    \item[\it \textbf{DustELGs-Poz1}:] variable flux threshold applied to \DustMod\ to match the number density of Pozzetti's model \#1 at each redshift,
    \item[\it \textbf{DustELGs-Poz3}:] variable flux threshold applied to \DustMod\ to match the number density of Pozzetti's model \#3 at each redshift,
\end{enumerate}

\noindent
where we note that all the ELGs are, by construction, a subset of the full \sage\ catalogue used in the previous section. Likewise, the four additional `\textit{ELGs-Poz}' catalogues are sub-sets of the public \RawMod\ and \DustMod, respectively. \\

Instead of introducing a redshift-dependent flux cut -- which might be considered counter-intuitive, as Euclid will have a fixed flux threshold -- we could have also taken the \RawMod\ model as the starting point and tuned our dust extinction parameters until we match the P16 $dN/dz$ values, akin to what \citet{Zhai_2019} have done. But finding the best possible dust model is beyond the scope of this work and hence we prefer to adhere to the former approach. The main idea here is to restrict the model ELGs to the brightest ones that are still observable. And we have seen in \Fig{fig:LF} that applying the dust model basically just shifts the LF towards lower luminosities, especially at high redshift and for the brightest ELGs \citep[e.g.][]{Sobral2016}. Therefore, adjusting the luminosity threshold will still select the brightest galaxies. Moreover, one could also re-calibrate \sage, the \getemlines\ code or choose a different dust model beyond a Cardelli law, all of which can affect the number density of ELGs. But exploring all these possibilities is beyond the scope of the present work. We prefer to work with minimal variations to the existing models and codes.

We also like to emphasize that our `\textit{-Poz1}' and `\textit{-Poz3}' models are \textit{not} the two models \#1 and \#3 of P16. They are ELG catalogues where we adjusted the number densities to match those of model \#1 and \#3 of Pozzetti, respectively. We did this to correct for the mismatch of ELGs with respect to the Pozzetti models seen in \Fig{fig:dNdz}. We further refrain from showing their abundance evolution as they match -- by construction -- the curves from P16.\\ 


Given the results presented in this section, we conclude that our UNITSIM-\sage-ELGs provide a fair sample and can be used for further analysis. The \RawMod\ and \DustMod\ galaxies will serve as the two base catalogues, with the four additional catalogues acting as our best predictions for Euclid. As a particular application we employ them now for a study of galaxy clustering and the related bias.

\section{Clustering of ELGs}\label{sec:corr}
Quantifying the clustering of galaxies is one of the main objectives of ongoing and upcoming galaxy surveys such as the Euclid satellite mission.  Clustering measurements probe the fluctuations of the underlying dark matter from the positions of galaxies, and they encode geometric, model-dependent cosmological information. Using the positions of our theoretical UNITSIM ELGs, we now study the two-point correlation function $\xi_{\rm ELGs}(r)$ and its redshift evolution. We further use the positions of $10^7$ randomly selected dark matter particles from the total $4096^3$ particles present in each of the UNIT gravity-only simulations to calculate $\xi_{\rm DM}(r)$.\footnote{We confirm that the resulting 2PCFs have converged and will not change when using more particles. Further, this number of dark matter particles is comparable to the number of ELGs, at least at redshifts $z\leq1$.} This allows us to also infer the bias that we define here as
\begin{equation} \label{eq:bias}
    b(r) = \sqrt{\frac{\xi_{\rm ELGs}(r)}{\xi_{\rm DM}(r)}}
\end{equation}
between both populations and study its evolution across redshift. The bias $b$, i.e. the statistical relation between the distribution of galaxies and matter, needs to be taken into account when interpreting galaxy surveys; it describes how galaxies trace the underlying dark matter distribution. The biased galaxy formation scenario \citep[e.g.][]{Dekel1987} implies that galaxies are not uniformly distributed in the Universe, but primarily form in the peaks of the matter density field. Galaxies are therefore biased tracers of it, sampling only the overdense regions \citep[see][for a recent review]{Desjacques2018}. The particular bias of ELGs, i.e. a sub-class of all galaxies, will be of greatest importance for surveys such as Euclid.

All two-point correlation functions (2PCFs) have been obtained with the \textsc{CUTE}\footnote{\url{https://github.com/damonge/CUTE}} software presented in~\citet{alonso2012cute}. In addition, for the results that we will present throughout this section, we have taken the average of the values computed for the 2PCF over the four simulations UNITSIM1, UNITSIM1-InvertedPhase, UNITSIM2, and UNITSIM2-InvertedPhase. 

\begin{figure}
   \includegraphics[width=\columnwidth]{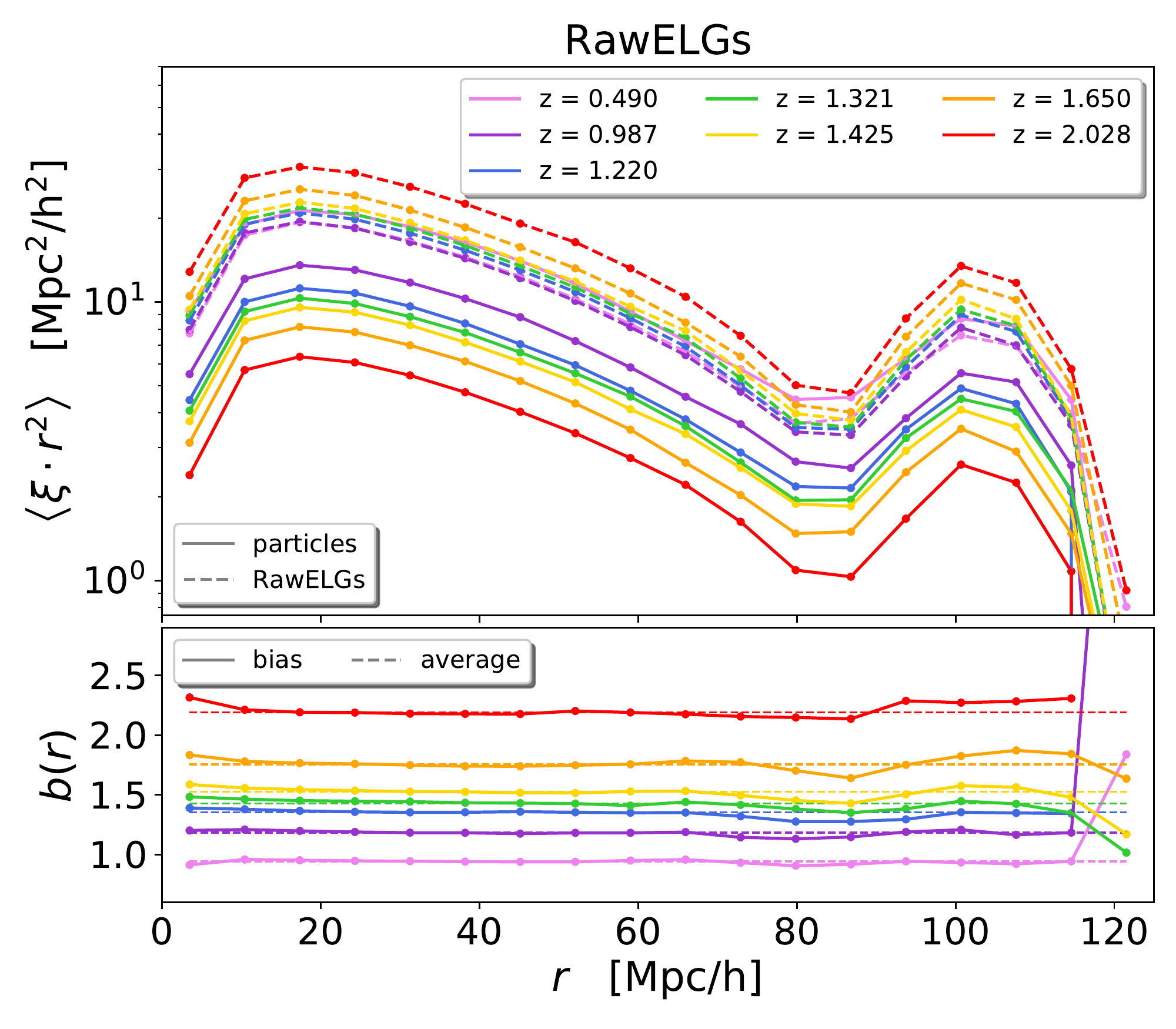}\vspace*{-0.2cm}
   \caption{Top panel: 2PCF of \sage\ ELGs with flux greater than $2\times 10^{-16}$~erg s$^{-1}$ cm$^{-2}$ (\RawMod\ galaxies, dashed lines) and collisionless trace particles (solid lines) for various redshifts. Bottom panel: associated bias as defined by \Eq{eq:bias}.}
 \label{fig:2PCF_RawMod}
 \vspace{-0.3cm}
 \end{figure}

In the top panel of \Fig{fig:2PCF_RawMod} we present the 2PCF computed for the \RawMod\ (dashed lines) and dark matter (solid lines). The lower panel of the same figure shows the bias $b(r)$ defined via \Eq{eq:bias}. In order to better verify the scale-dependence of the bias, we also calculate the `average' bias
\begin{equation} \label{eq:meanbias}
    \langle b\rangle = \frac{1}{N_{\rm bin}-1} \sum_{2}^{N_{\rm bin}} b_i \ ,
\end{equation}
where $N_{\rm bin}$ is the number of bins and $b_i=b(r_i)$ the value of the bias in distance bin $r_i$. This average bias $\langle b\rangle$ is shown as a dashed horizontal line in the lower panel of \Fig{fig:2PCF_RawMod}. Note that we exclude the first bin in this calculation since for such small distances the bias is certainly scale-dependent (see \Fig{fig:bias_SSS} below). It is also obvious that the data for this particular model become rather noisy at high redshifts due to the very low number of objects above the reference flux cut of $F_{\rm cut}=2\times 10^{-16}$~erg s$^{-1}$ cm$^{-2}$ (see \Tab{tab:FluxThresholds}). But we can nevertheless appreciate that for distances $r\gsim 5\hMpc$ the bias is remarkably constant, something we will quantify in more detail below.

\begin{figure}
   \includegraphics[width=\columnwidth]{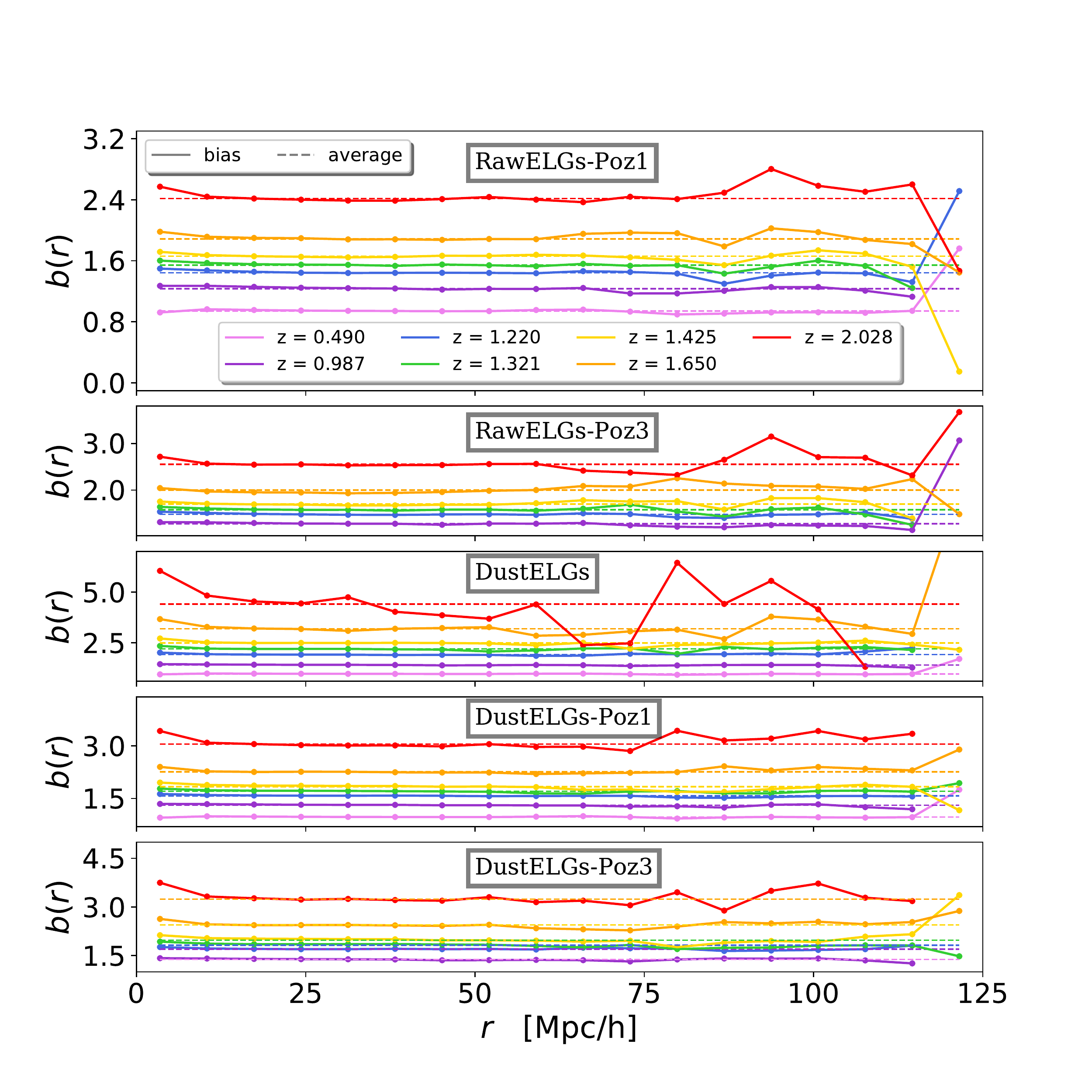}\vspace*{-0.2cm}
   \caption{The bias for the \PozModAraw, \PozModCraw, \DustMod, \PozModAdust, and \PozModCdust\ galaxies (in that order from top to bottom), using the same $r$-range and redshift colouring as for \Fig{fig:2PCF_RawMod}.}
 \label{fig:bias_LSS}
 \vspace{-0.3cm}
 \end{figure}

An equivalent analysis has been conducted for our other ELG catalogues, but we decided to only show here in \Fig{fig:bias_LSS} the results for the bias and not also the 2PCFs. Once more we can see that we get fairly noisy results at redshift $z = 2.028$ due to the reduced number of galaxies at that redshift. We also observe that at scales $\sim 120\hMpc$ the bias behaves more erratic, which can be explained by the fact that the 2PCF crosses zero at $r\sim 130\hMpc$ \citep[][]{Sanchez2008,Prada2011arxiv}: taking the numerical ratio between two numbers close to zero then introduces noise. But the most important point is that the bias of ELGs (at least for $z<2$) in all our catalogues remains constant on scales $r\in [5,100]\hMpc$ \citep[in line with the findings of, for instance,][]{Abbott2018}. Below 5\hMpc\ it is obvious that the mixture contribution between the one- and two-halo terms will introduce non-linear effects which in turn will cause the bias to no longer behave independently with scale. On larger scale we have already seen above that the zero-crossing of the 2PCF is introducing noise and hence the results for the bias are expected to be affected by this, too. We further note that the bias clearly is a function of redshift. But this is also expected, as the mass of the haloes hosting ELGs will change with redshift (see \Fig{fig:HMFofELGs} in the Appendix). Not only that, but haloes of the same mass or luminosity at different redshifts will also have a different bias. It therefore only appears natural that the bias increases with redshift as, for instance, modelled analytically by \citet{Basilakos2008} or found in other cosmological simulations \citep[e.g.][]{Merson2019, Tutusaus2020}.

  \begin{figure}
   \includegraphics[width=\columnwidth]{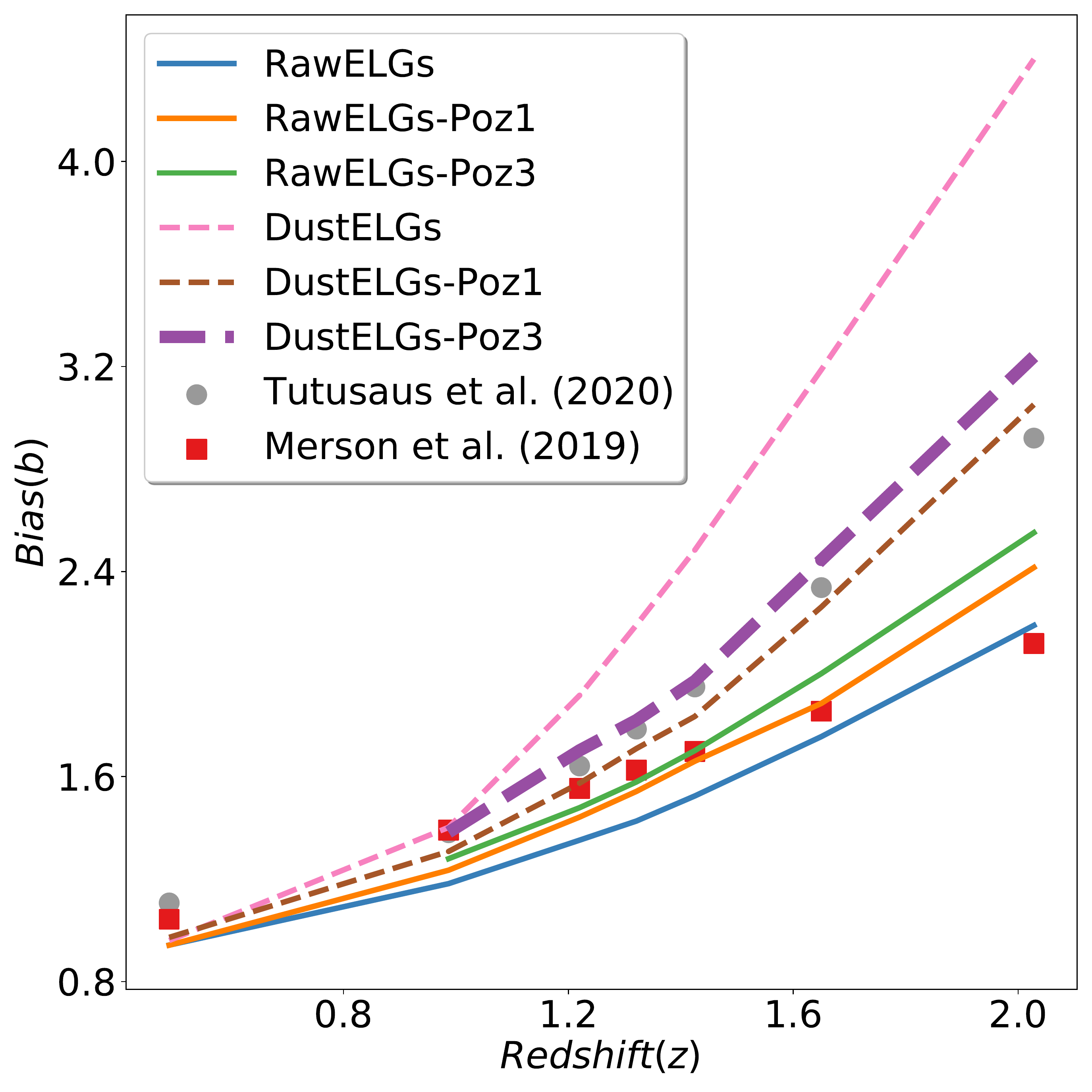}\vspace*{-0.2cm}
   \caption{Bias values averaged for scales larger than 5 \hMpc\ computed for our six models. The grey points are the best fit $b(z)$ for the bias found in Euclid's \textit{Flagship} simulation \citep{Tutusaus2020}, and the red squares the results as reported by \citet{Merson2019}.}
 \label{fig:bias_vs_z}
 \vspace{-0.3cm}
 \end{figure}

\begin{table*}
\caption{Bias values averaged for scales larger than 5\hMpc\ computed for all our ELG catalogues. The values listed here correspond to the lines presented in \Fig{fig:bias_vs_z}.}
\begin{center}
\begin{tabular}{ccccccc}
\hline
$z$   & \RawMod & \PozModAraw & \PozModCraw & \DustMod & \PozModAdust & \PozModCdust \\
\hline
\hline
0.490 & 0.94 & 0.94 & --   & 0.96   & 0.97    & --      \\
0.987 & 1.18 & 1.23 & 1.28 & 1.40   & 1.31    & 1.38    \\
1.220 & 1.35 & 1.44 & 1.48 & 1.92   & 1.57    & 1.70    \\
1.321 & 1.42 & 1.54 & 1.58 & 2.19   & 1.71    & 1.82    \\
1.425 & 1.53 & 1.66 & 1.70 & 2.48   & 1.83    & 1.97    \\
1.650 & 1.76 & 1.89 & 2.00 & 3.19   & 2.26    & 2.45    \\
2.028 & 2.19 & 2.42 & 2.55 & 4.40   & 3.05    & 3.24    \\ \hline
\end{tabular}
\label{tab:bias}
\end{center}
\end{table*}

\Fig{fig:bias_vs_z} now quantifies the evolution of the average bias $\langle b\rangle$ (obtained from the results presented in \Fig{fig:2PCF_RawMod} and \Fig{fig:bias_LSS}) as a function of redshift for all our catalogues. This figure is accompanied by \Tab{tab:bias} that lists the plotted values. We find that for all our galaxies the bias systematically increases with redshift, despite showing different growth rates, especially for the two base catalogues \RawMod\ and \DustMod. We also acknowledge that the strength of this $b(z)$ relation for our four `\textit{-Poz}' galaxies -- especially the ones based upon \DustMod\ -- is in excellent agreement with the relation presented in \citet[][eq.~11]{Tutusaus2020}, shown as circles in \Fig{fig:bias_vs_z}. The $b(z)$ function given in Tutusaus et al. is derived from studying the bias in the Euclid \textit{Flagship} simulation,\footnote{\url{https://www.euclid-ec.org/?page_id=4133}} which is also just based upon dark matter. But the way in which the dark matter haloes are populated with galaxies is quite distinct to our approach: they have applied a Halo Occupation Distribution (HOD) that does not take into account the merger trees of the haloes \citep[for a comparison of these two different techniques see, for instance,][]{Knebe15,Knebe2018}.\footnote{While there is no reference paper for this galaxy catalogue, we nevertheless like to mention that it is based upon the MICE HOD \citep{Carretero14}. The clustering is fit to SDSS galaxies as a function of magnitude and colour at low redshift. Then, most of the properties are assumed to depend on redshift only via their SEDs/color evolution, allowing for correlations between many observables.}
\citet{Merson2019} also forecast the redshift evolution of the linear bias for \Halpha-emitting galaxies in a similar redshift range. Their data are shown here as squares. Like \citet{Tutusaus2020}, they also used a HOD for which they calibrated the dust attenuation to reproduce observed \Halpha\ counts. \citet{Merson2019} now predict lower biases than \citet{Tutusaus2020} and our dust-based `\textit{-Poz}' galaxies, more in line with the results we obtain for our \RawMod\ catalogue and its derivates. The comparison of these three different $b(z)$ predictions for ELGs indicates that the theoretical models have not yet converged. There are degeneracies and uncertainties that still require more detailed and refined investigations before any final conclusion could be drawn. But we finally remark that our findings for the redshift evolution of the bias $b(z)$ are also in agreement with those of \citet[][right panel of their fig.~6]{Favole17}, who used a SHAM model. However, in their work, the bias increases more mildly, as the SDSS redshift range studied there is very much reduced compared to ours.

We further recognize in \Fig{fig:bias_vs_z} that the bias is sensitive to the particulars of our modelling, especially at high redshift. This certainly relates to how we treat the dust extinction and select the observable ELGs from \RawMod, respectively. But this is known and can also be appreciated when comparing the bias predictions from \citet{Tutusaus2020} and \citet{Merson2019} where similar discrepancies are seen. We particularly notice the degeneracy between dust modelling and flux selection: first applying our extinction prescription and then matching a preset $dN/dz$ by varying the flux threshold always leads to larger bias than not employing a dust model at all. Even though we argued before that the dust-attenuated luminosities -- as seen in \Fig{fig:LF} -- are a shifted version of the raw values \citep[at least for luminous ELGs; see also][where a constant luminosity offset was applied to model dust extinction]{Sobral2016}, here we realize that their relation is not that simple. But we have clearly seen that fixing the abundance of \Halpha\ ELGs, the differences substantially reduce. Nevertheless, we like to stress again that designing a new dust extinction model is beyond the scope of this work and hence we leave a more detailed study of this to a future work. Note that in this work we primarily aim at presenting the publicly available data, discussing its scope and possible limitations.\\

So far we have mainly focused on large scales, but to conclude this section we also present how the bias varies for \textit{small} scales. In \Fig{fig:bias_SSS} we present the bias $b(z)$ for various redshifts and all our catalogues out to $r\approx20\hMpc$ using logarithmic binning. We observe that for redshifts $z<2$ the bias remains constant down to scales $r\approx 3\hMpc$ and then starts to mildly drop. It is actually around this distance that we expect the contribution from the one-halo term to start to become relevant. However, this behaviour weakens for higher redshifts and possibly reverses for $z=2$. Something similar has also been observed by \citet[][fig.~10]{Nuza13} for BOSS CMASS galaxies, but there the inversion was already seen at redshift $z\approx0.53$ (and one needs to bear in mind that CMASS galaxies and ELGs are not directly comparable as they are different types of galaxies, where the latter are mostly star-forming and the former could be dominated by passive galaxies).

  \begin{figure}
   \includegraphics[width=\columnwidth]{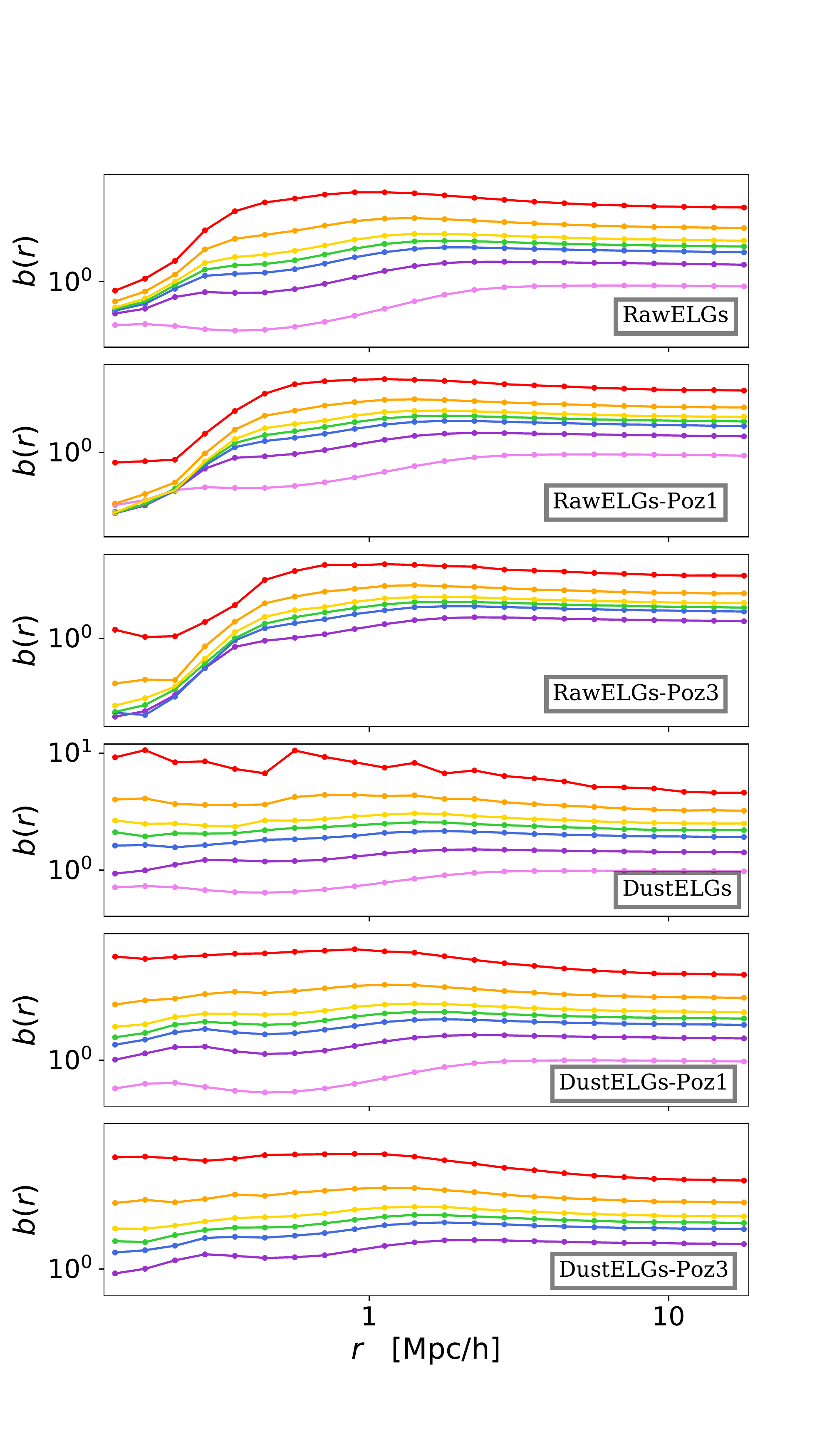}\vspace*{-0.2cm}
   \caption{The bias $b(r)$ for all six models in logarithmic $r$ bins, focusing on the small scales up to $r<20$\hMpc, using the same redshift colouring as for \Fig{fig:2PCF_RawMod}.}
 \label{fig:bias_SSS}
 \vspace{-0.3cm}
 \end{figure}

\section{Conclusions} \label{sec:conclusions}
Realistic simulations are a necessary tool to optimise and validate the methodology that will be used to extract cosmological constraints from future surveys. Indeed, they are used to estimate the theoretical error budget on surveys \citep[for example, for the eBOSS-ELG analysis, see][]{Alam2020}. 
In this work we have employed the UNIT simulations, which model the evolution of dark matter within a 1\hGpc\ box at a mass resolution of $1.2\times10^{9}\hMsun$ per particle \citep[][]{Chuang_2019}. Given the large volume of these simulations together with the fixed-and-paired technique of \citet[][]{Angulo_2016} that enhances the effective volume of the simulations, the resulting galaxy mocks that we have produced represent a unique resource for model testing based on a semi-analytic model of galaxy formation. We used our ELG catalogues to make predictions for the galaxy statistics that the Euclid experiment is expected to obtain for redshifts between $0.9<z<1.8$. Note that the simulations presented here cover an effective survey volume of about seven times the effective survey volume of Euclid \citep{Chuang_2019}. And having the galactic physics included is key, since the complicated relation between haloes and galaxies can modify the clustering of ELGs significantly, even at scales used to put cosmological constraints when working in Fourier space \citep[see, for instance, ][]{Gonzalez-Perez2020,Avila2020}.

For this work we have generated six synthetic catalogues of emission-line galaxies of which the two base ones (i.e. \RawMod\ and \DustMod, without any flux cuts applied) are publicly available. The galaxies were first obtained by applying the semi-analytic galaxy formation model \sage\ \citep[][]{Croton16} to the gravity-only UNIT simulations. They were then subjected to the emission line modelling with the \getemlines\ code \citep[][]{Orsi_2014} and an additional dust attenuation model \citep[following][]{Favole2020}. This left us with the two base ELG catalogues \RawMod\ and \DustMod, in addition to the general \sage\ galaxy catalogues. As argued throughout \Sec{sec:galaxies}, the properties associated with our UNITSIM-\sage\ galaxies reproduce observed properties of galaxies with $0\le z < 2$. Here we have focused on those properties that are most relevant for the construction of ELGs catalogues, i.e. stellar mass, star formation rate, metalicity, and disc size. In particular, we find that the (evolution of the) mass--metallicity relation agrees sufficiently well with observations. However, we have seen in \citet{Knebe2018b} that the \sage\ model underpredicts the number of galaxies with high SFRs. This then affects the abundance of our (dust-attenuated) ELGs as seen in \Fig{fig:dNdz}. While we presented the validation plots in the main body of the paper only for the full set of \sage\ galaxies, the corresponding plots for the \RawMod\ and \DustMod\ ELGs can be found in \App{app:ELGs}.

In \Sec{sec:ELGs} we adjusted the number densities of our two base UNITSIM-ELG sets by applying distinct flux thresholds to them (using the Euclid-models as given by P16), eventually comparing the redshift evolution of their abundance to observations. When studying the density of galaxies per $deg^2$ with fluxes greater than $2\times 10^{-16}$~erg s$^{-1}$ cm$^{-2}$ as a function of redshift we observe that the density obtained for the raw ELG galaxies is above both the observations and other theoretical modelling. That means that some additional selection needs to be applied to end up with a more realistic ELG catalogue. We have addressed this in several ways. We first applied a dust-attenuation \citep[a Cardelli law, following][]{Favole2020}, which led to a possible underestimation of the expected density of galaxies $dN/dz$ observed from redshift $z\sim 1.4$ onwards. Nevertheless, the most recent study by \citet{Bagley2020} suggests that the observational value for $dN/dz$ could be closer to our results than predicted by P16. We also designed additional catalogues where we instead varied the flux threshold for the selection of galaxies from the \RawMod\ catalogue; those fluxes were adjusted to reproduce number densities as predicted by P16.\\

The linear bias is a key parameter to understand the cosmological power of Euclid and can help construct forecasts that inform the optimisation of both observational and analysis strategies. The bias of \Halpha\ galaxies may be particularly relevant for forecasts on studies such as primordial non-Gaussianities or relativistic effects. We therefore studied the clustering of all our six samples listed in \Tab{tab:FluxThresholds}: two with the Euclid flux cut applied and four in which the flux cuts are adjusted to follow the predictions by two of the models presented in P16. We measure the linear bias as a function of redshift by averaging $\xi_{\rm ELGs}/\xi_{\rm DM}$ for scales $r>5\hMpc$. For the samples whose abundances are matched to the to P16 predictions, we find a $b(z)$ in line with that reported in \citet{Tutusaus2020} for the Euclid \textit{Flagship} simulation \citep[and mildly in agreement with the same results reported by][]{Merson2019}. This is striking, as the \textit{Flagship} mock was constructed following a very different methodology \citep{Carretero14}. Additionally, we report the clustering at small scales, that becomes scale-dependent. These measurements can be used to test the robustness of different large-scale structure models to extract cosmological information from the small scales, that have the highest signal-to-noise ratio but at the same time are the most difficult to model.\\

We close with the remark that an improved dust attenuation modelling might be the most physical approach for choosing the ELGs so that the observed $dN/dz$ will be recovered. This would, however, only affect the catalogues that are based upon \DustMod; it will leave \RawMod\ untouched, which is the primary ELG catalogue made available publicly. Therefore, while we have shown throughout this work that the particulars of the dust extinction have an effect on the the results, the published data contain all that is required for the community to apply their favourite post-processing models for dust and emission lines from star-forming regions. Or put differently, the base catalogue \RawMod\ is certainly inclusive, i.e. a superset of the ELGs of interest. A better understanding of the process of selecting observable ELGs from that base catalogue and developing an improved dust attenuation model will be left for a future work. The public data can already been used for a great variety of studies and have extensive applications like, for example, informing Halo Occupation Distribution models. Indeed, we will study the properties of \Halpha\ ELG HOD models in a follow-up paper. \\

\section*{Acknowledgements}
\addcontentsline{toc}{section}{Acknowledgements}
We thank the anonymous referee for a detailed and constructive report that helped to improve the content of the paper and its presentation.

AK and GY are supported by the MICIU/FEDER through grant number PGC2018-094975-C21. AK further thanks Chasing Dorothea for all i want. SA is supported by the MICUES project, funded by the EU H2020 Marie Skłodowska-Curie Actions grant agreement no. 713366 (InterTalentum UAM). DLC acknowledge the support of the ERC-StG number 716151 (BACCO). GF acknowledges financial support from the SNF 175751
“Cosmology with 3D Maps of the Universe” research grant. VGP and SA are supported by the Atracción de Talento Contract no. 2019-T1/TIC-12702 granted by the Comunidad de Madrid in Spain. ARHS acknowledges receipt of the Jim Buckee Fellowship at ICRAR-UWA. FSK acknowledges financial support from the Spanish Ministry of Economy and Competitiveness (MINECO) under the Severo Ochoa program SEV-2015-0548, and for the grants RYC2015-18693 and AYA2017-89891-P. GR is grateful for the financial support granted by the Ministry of Education of Spain (FPI, MEC) with reference PRE2018-087035, for the I+D project with reference to SEV -2016-0597 - 18-2 IFT (UAM-CSIC).

The UNIT simulations have been produced in  the MareNostrum Supercomputer,  hosted by the Barcelona Supercomputing Center, Spain, under  the PRACE project number 2016163937.

This research has made use of NASA’s Astrophysics Data System and the arXiv preprint server. This work was created by making use of the following software: \textsc{Python}, \textsc{Matplotlib} \citep{Matplotlib-Hunter07}, \textsc{Numpy} \citep{Numpy-vanDerWalt11}, \textsc{scipy} \citep{Scipy-Virtanen19}, and \textsc{astropy} \citep{AstropyI,AstropyII}.


\section*{Data Availability}
The data are available at \url{http://www.unitsims.org}. There one finds the halo catalogues and merger trees of the underlying dark matter--only simulations as well as the \sage\ galaxies alongside their emission-line information. These latter ELG files contain the two base catalogues \RawMod\ and \DustMod, but without applying any flux cut.

All files are in hdf5 format. Sample reading routines that specify the properties stored inside the files (and their units) are available there too, but can also be downloaded from here:\\

\noindent
\url{http://popia.ft.uam.es/public/read_SAGE.py}\\
\url{http://popia.ft.uam.es/public/read_ELGs.py}\\

The data are directly available for those redshifts used in this paper; for other redshifts the data will be available upon request.


\bibliographystyle{mnras}
\bibliography{archive}


\appendix

\section{Additional validation plots}\label{app:additional}
In this Appendix we provide supplementary plots that further show the validity and properties of the \sage\ and ELG galaxies used throughout this study. 

\subsection{Halo Mass Function of flux-selected ELGs} \label{app:HMFofELGs}
\begin{figure}
   \includegraphics[width=0.95\columnwidth]{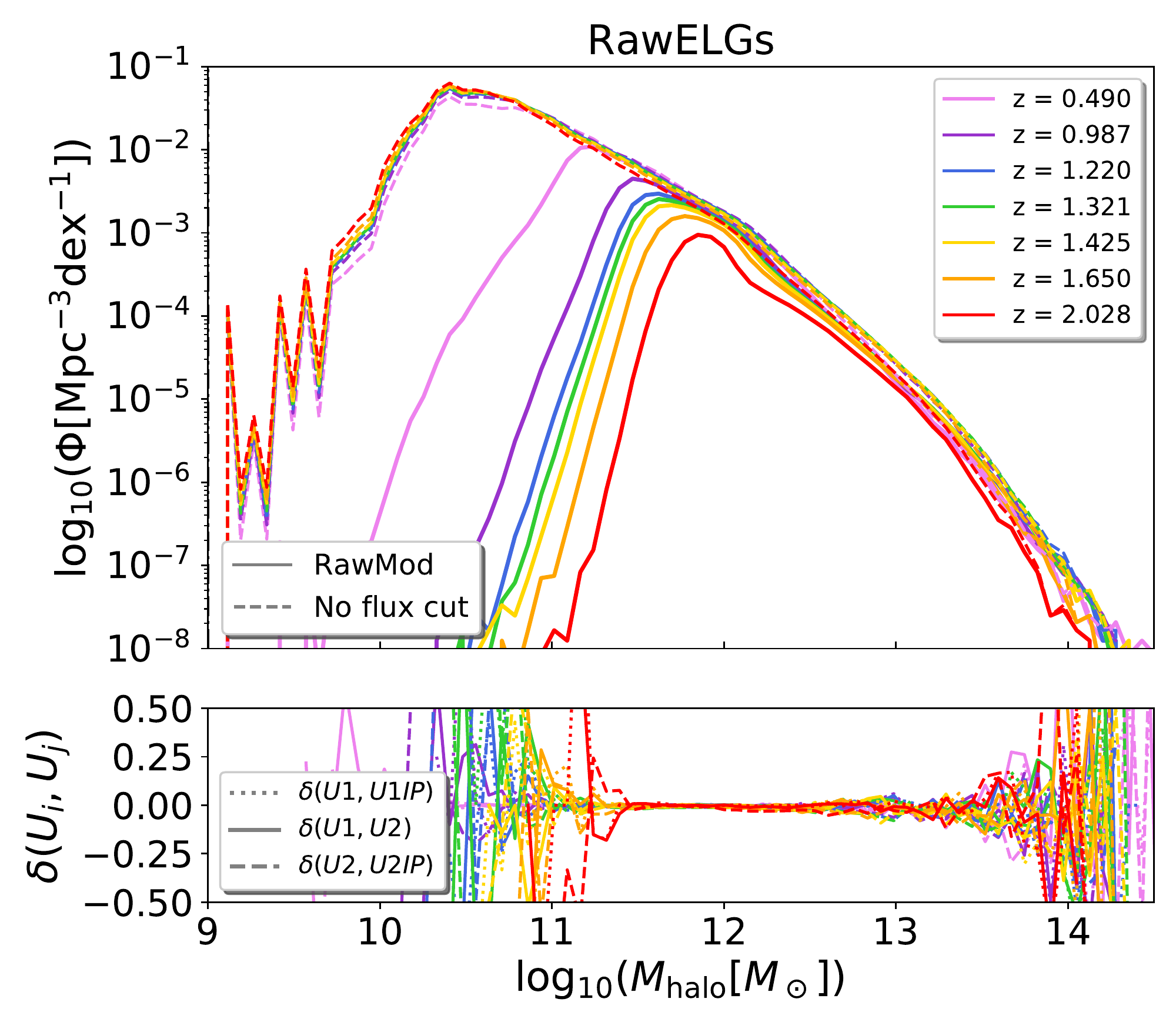}
   \includegraphics[width=0.95\columnwidth]{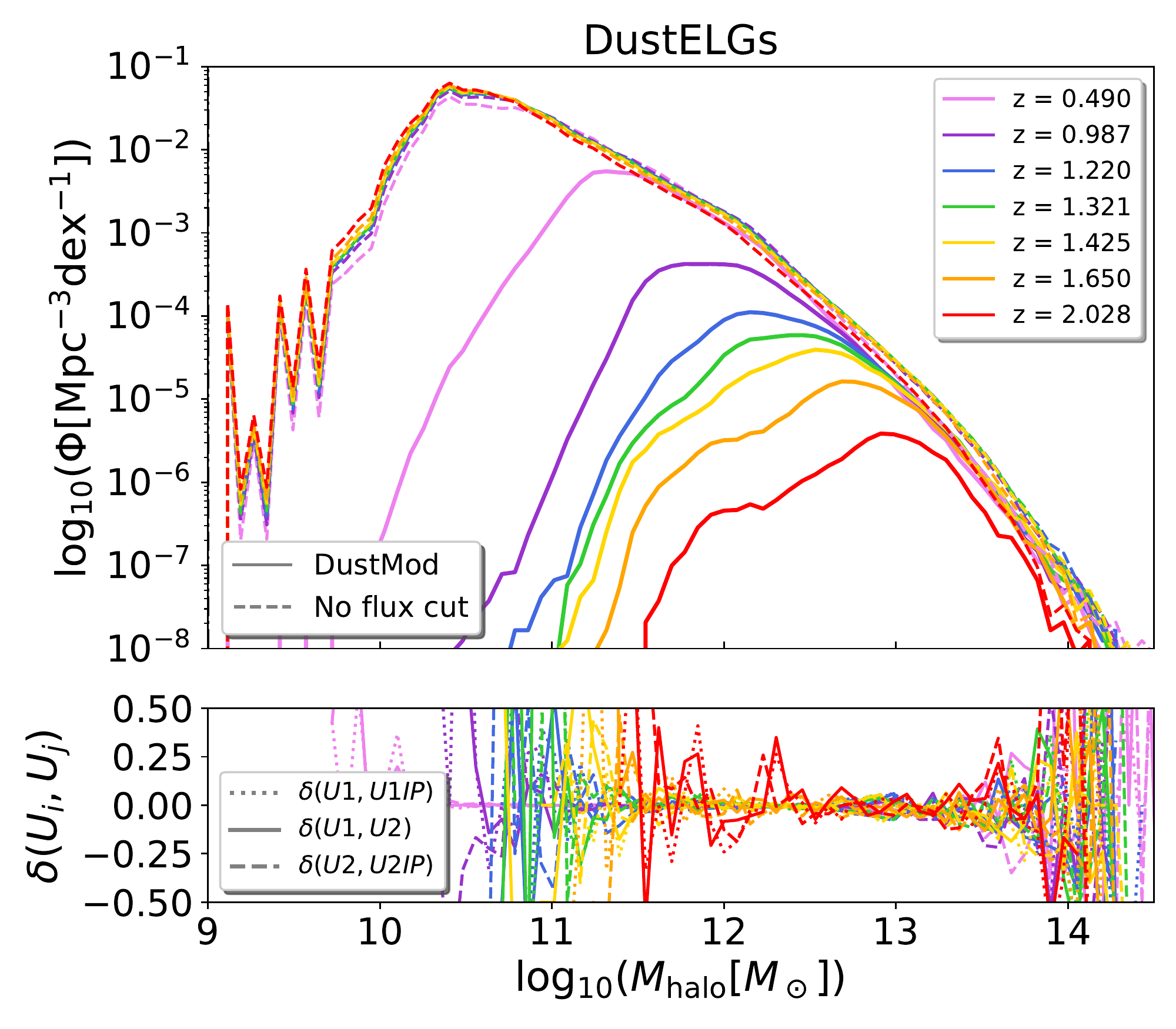}
   \caption{Halo mass function of all ELGs (dashed lines) and the flux-selected samples (solid lines) for \RawMod\ (top) and \DustMod\ (bottom).}
 \label{fig:HMFofELGs}
 \vspace{-0.3cm}
 \end{figure}

Applying a SAM will eventually lead to selecting a sub-sample of the underlying dark matter haloes as galaxies, i.e. while every halo contains a galaxy, some might be too small to be detectable. To better understand which haloes host our ELGs, we show their halo mass functions for the two base models \RawMod\ and \DustMod\ for various redshifts in \Fig{fig:HMFofELGs}. The dashed lines are without applying any flux cut, whereas the solid lines use the Euclid-inspired cut $F_{\rm cut}=2\times 10^{-16}$~erg s$^{-1}$ cm$^{-2}$. We can see that the flux cut primarily affects low-mass haloes, i.e. the less luminous ELGs also live in lower mass host haloes. We further observe a shift of this `cut-off' halo mass with redshift; while at $z\sim 0.5$ it is approximately $10^{11} M_{\odot}$, it increases to $\sim 10^{12} M_{\odot}$ at $z\sim2$ for \RawMod\ and even $\sim 10^{13} M_{\odot}$ for \DustMod.

\subsection{Baryonic properties of flux selected ELGs} \label{app:ELGs}
In \Sec{sec:galaxies} we presented baryonic relations for the full set of \sage\ galaxies, focusing on those properties that are relevant for the dust attenuation modelling. Here we now like to provide counterparts of those plots for the ELGs.

\subsubsection{Stellar Mass Function} \label{app:SMFofELGs}
\begin{figure}
   \includegraphics[width=0.95\columnwidth]{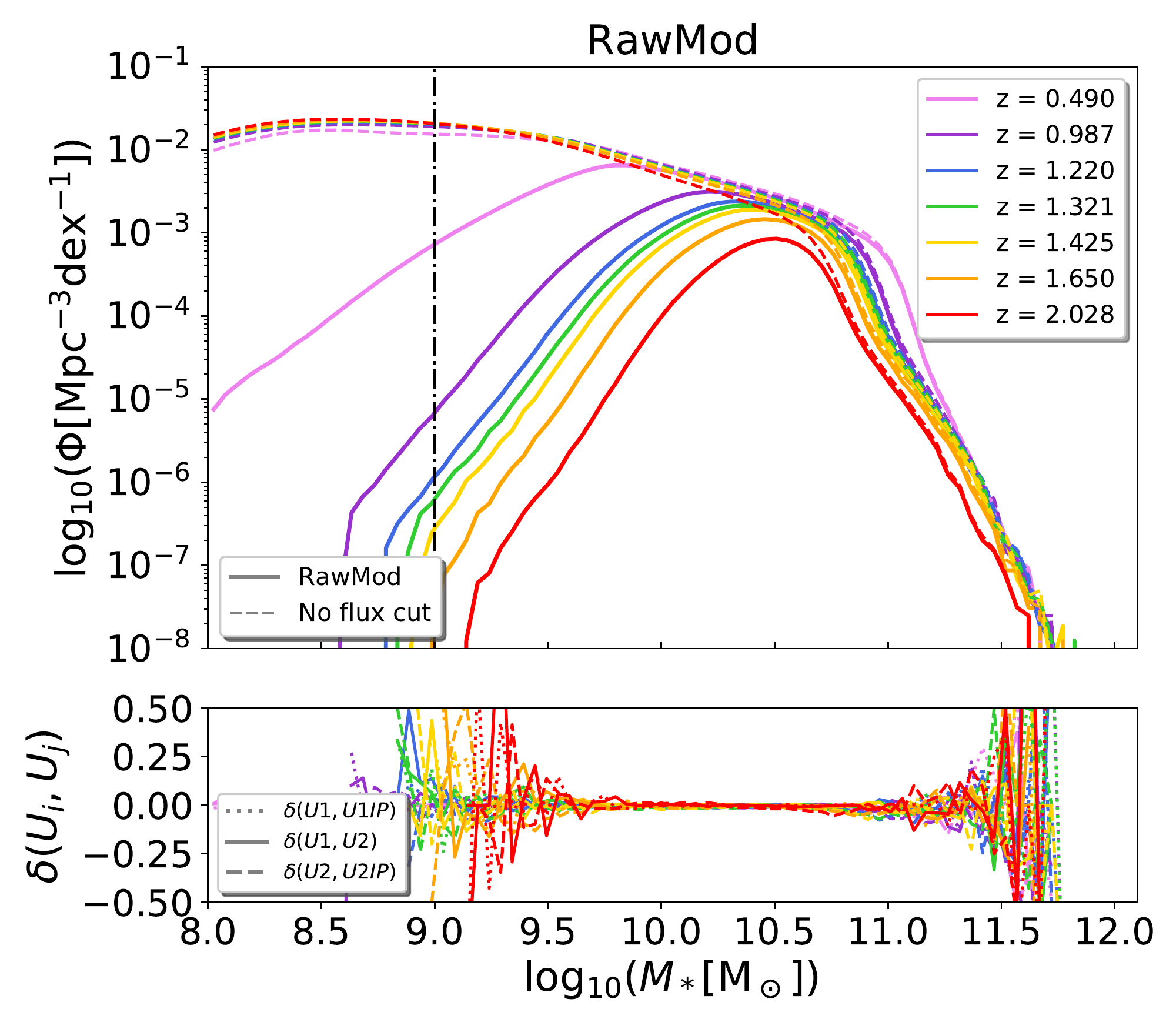}
   \includegraphics[width=0.95\columnwidth]{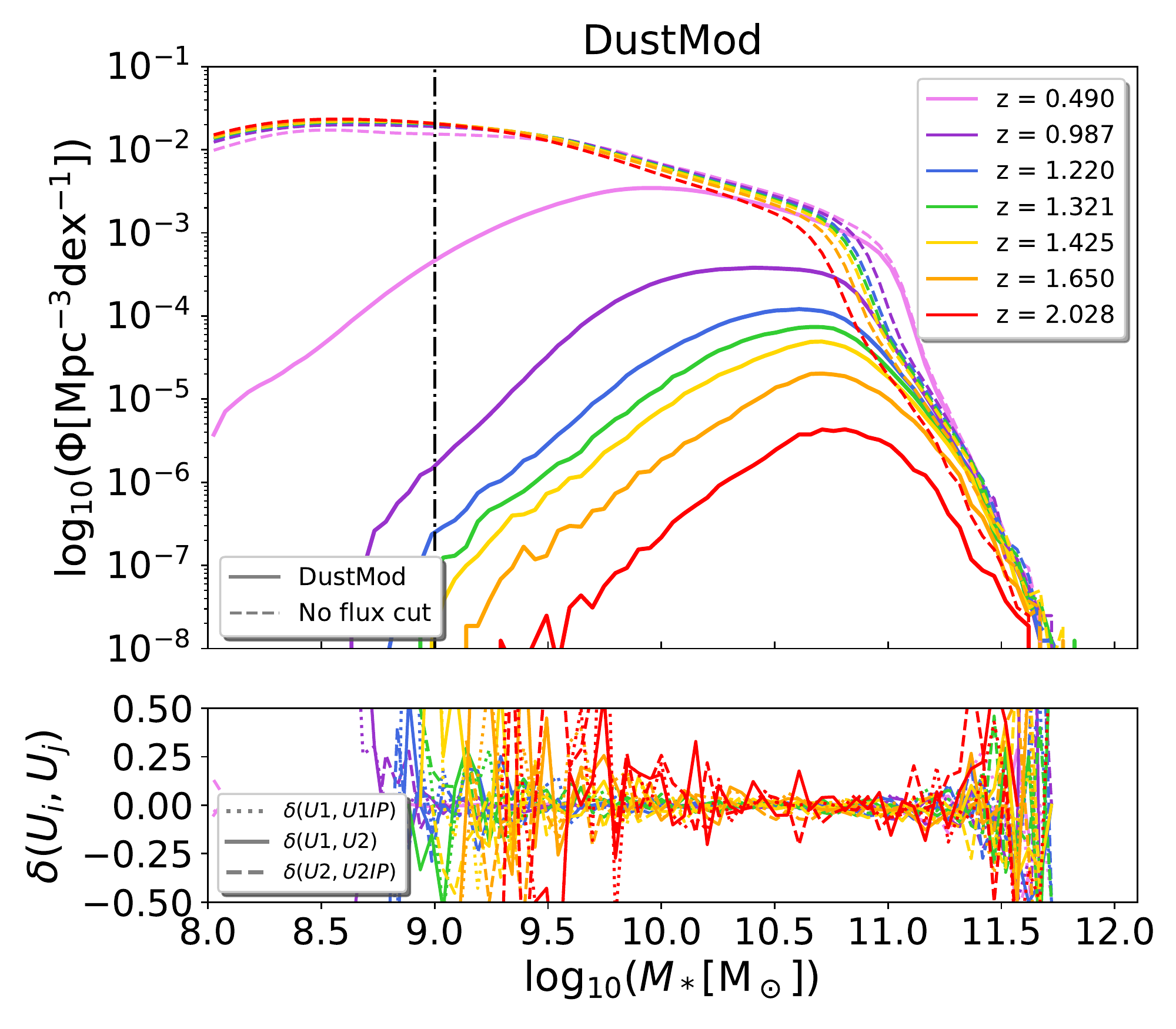}
   \caption{Stellar mass function of all ELGs (dashed lines) and the flux-selected samples (solid lines) for \RawMod\ (top) and \DustMod\ (bottom). The vertical dot-dashed line shows our lower stellar mass limit.}
 \label{fig:SMFofELGs}
 \vspace{-0.3cm}
 \end{figure}

In order to view the effect of the flux selection and its relation to the stellar masses of the resulting sub-sample of ELGs, we show in \Fig{fig:SMFofELGs} both the SMF of all ELGs (i.e. no flux cut, dashed lines) and the flux-selected samples of ELGs (solid lines) for various redshifts. We restrict the results again to the two base models \RawMod\ and \DustMod. We appreciate that the majority of ELGs coincide with the most massive galaxies.

\subsubsection{Specific star formation rate} \label{app:sSFRofELGs}
\begin{figure}
   \includegraphics[width=0.95\columnwidth]{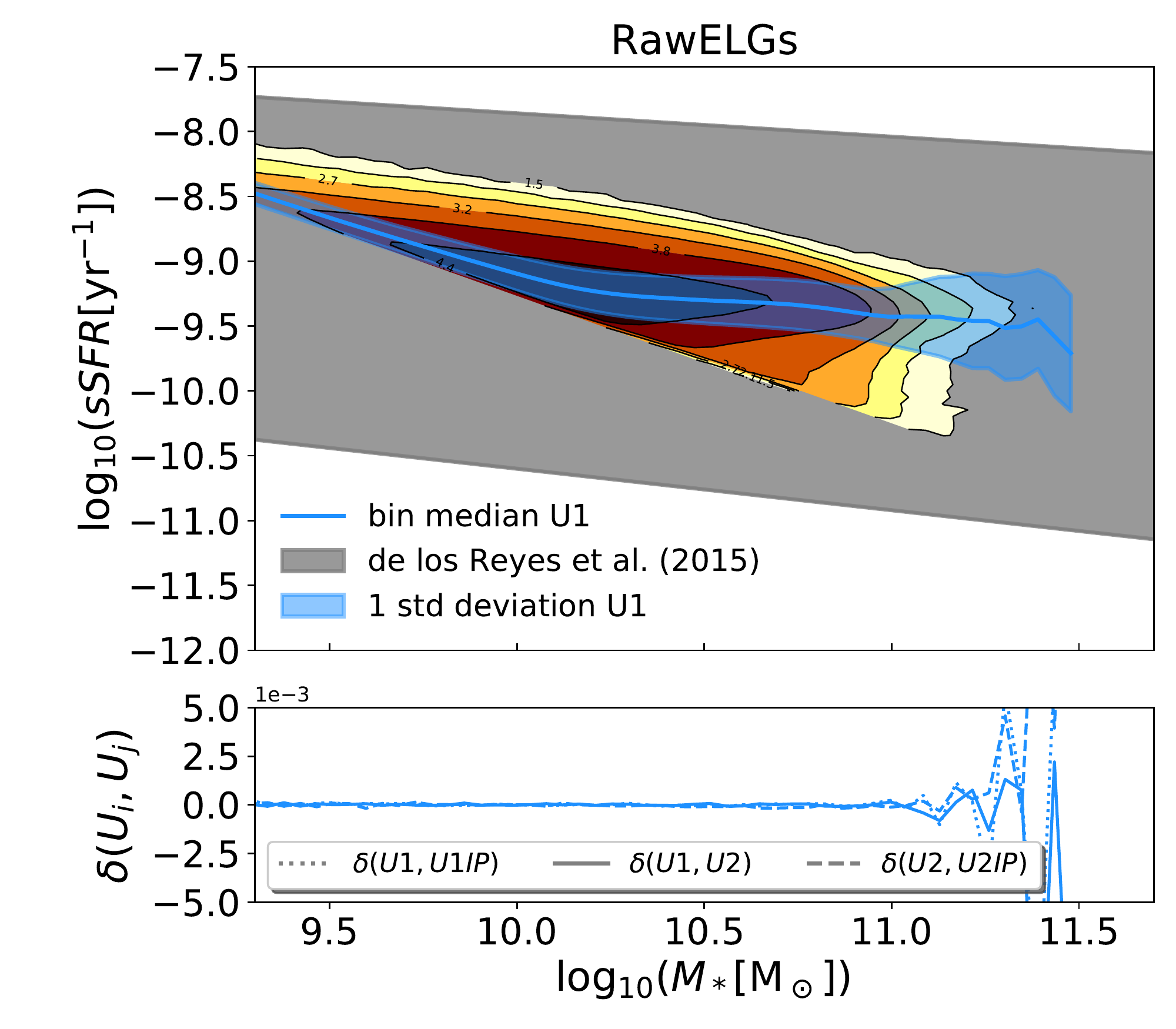}
   \includegraphics[width=0.95\columnwidth]{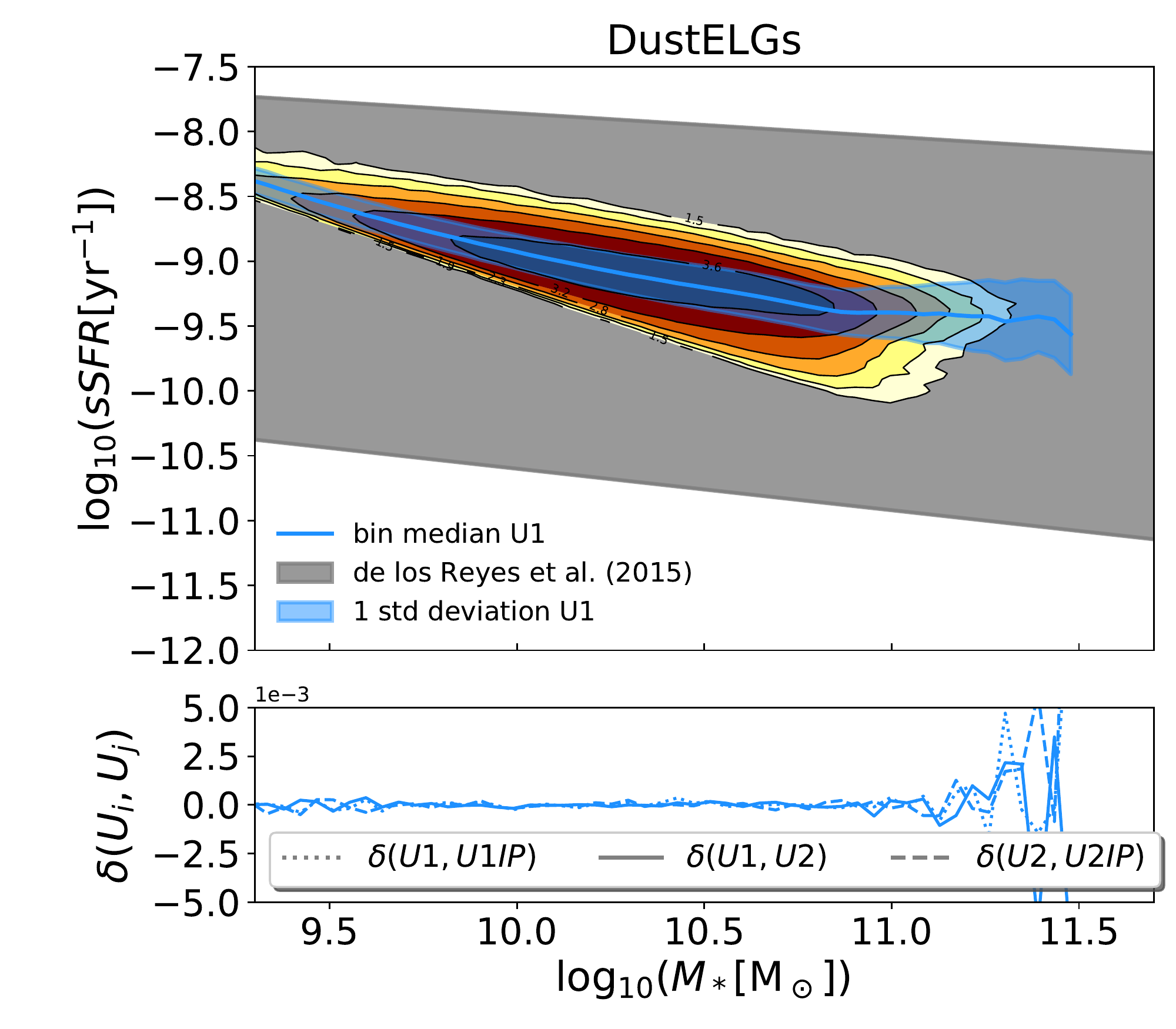}
   \caption{Specific star formation rate of the \RawMod\ (top) and \DustMod\ (bottom) ELGs at redshift $z\sim 1$ in comparison to the best-fit relation as found by \citet{delosReyes2015} at $z\sim 0.8$, shown as grey-shaded region. This figure is a reproduction of \Fig{fig:sSFR2028}, but this time for our model ELGs.}
 \label{fig:sSFRofELGs}
 \vspace{-0.3cm}
 \end{figure}

In \Fig{fig:sSFR2028} we show the specific star formation rate of all our \sage\ galaxies in comparison to the observations of \citet{Daddi2007} at redshift $z\sim2$. Here we now present in \Fig{fig:sSFRofELGs} another version of that plot, this time using the (flux-cut) ELGs of the \RawMod\ and \DustMod\ catalogues. We further show results for $z\sim 1$ and add the best-fitting correlation for \Halpha\ emitting galaxies, as found by \citet[][eq.~3]{delosReyes2015}.\footnote{\citet{delosReyes2015} studied 299 \Halpha-selected galaxies at redshift $z\sim 0.8$.}

\subsubsection{The mass--metallicity relation} \label{app:ZofELGs}
\begin{figure}
   \includegraphics[width=0.95\columnwidth]{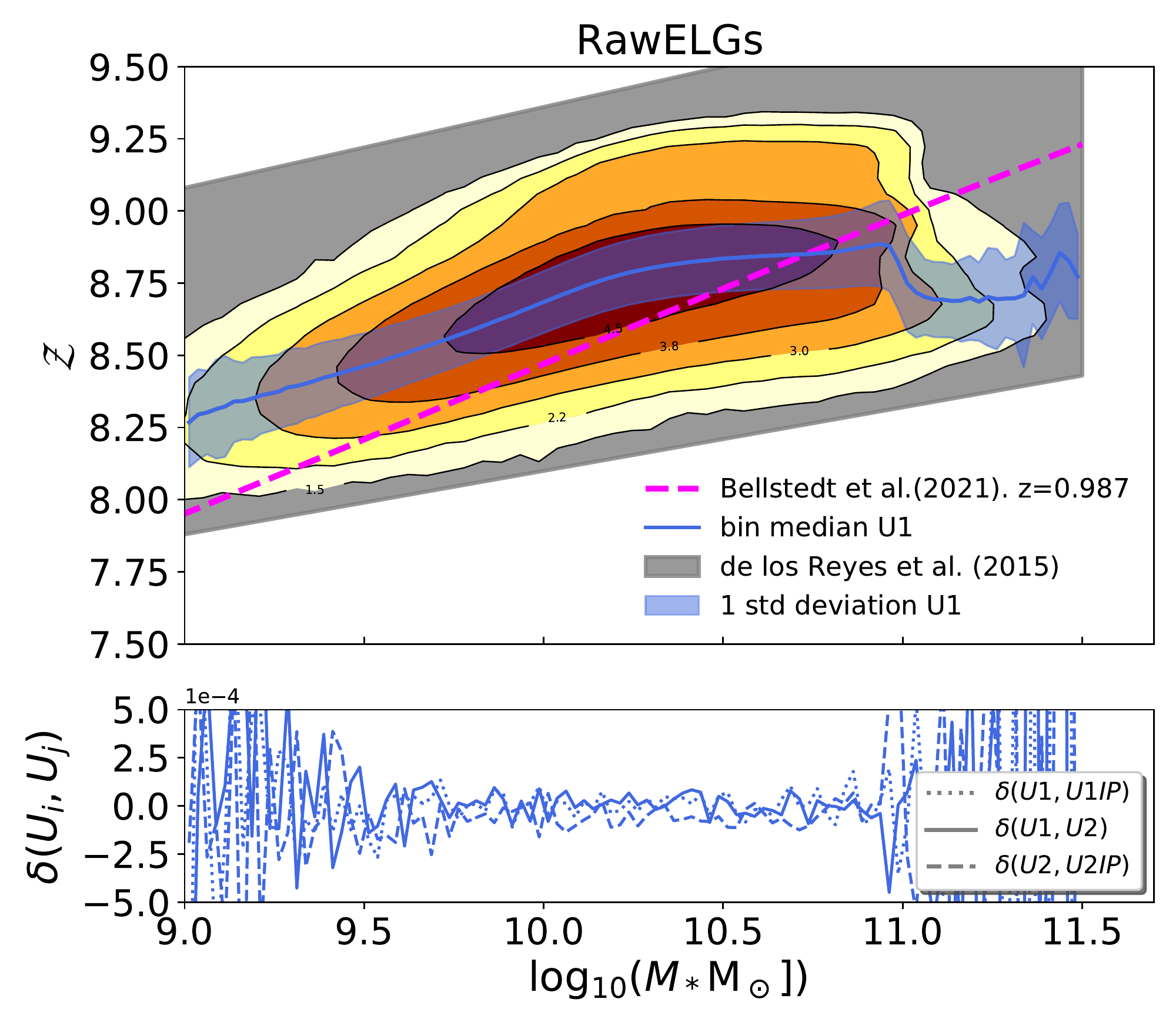}
   \includegraphics[width=0.95\columnwidth]{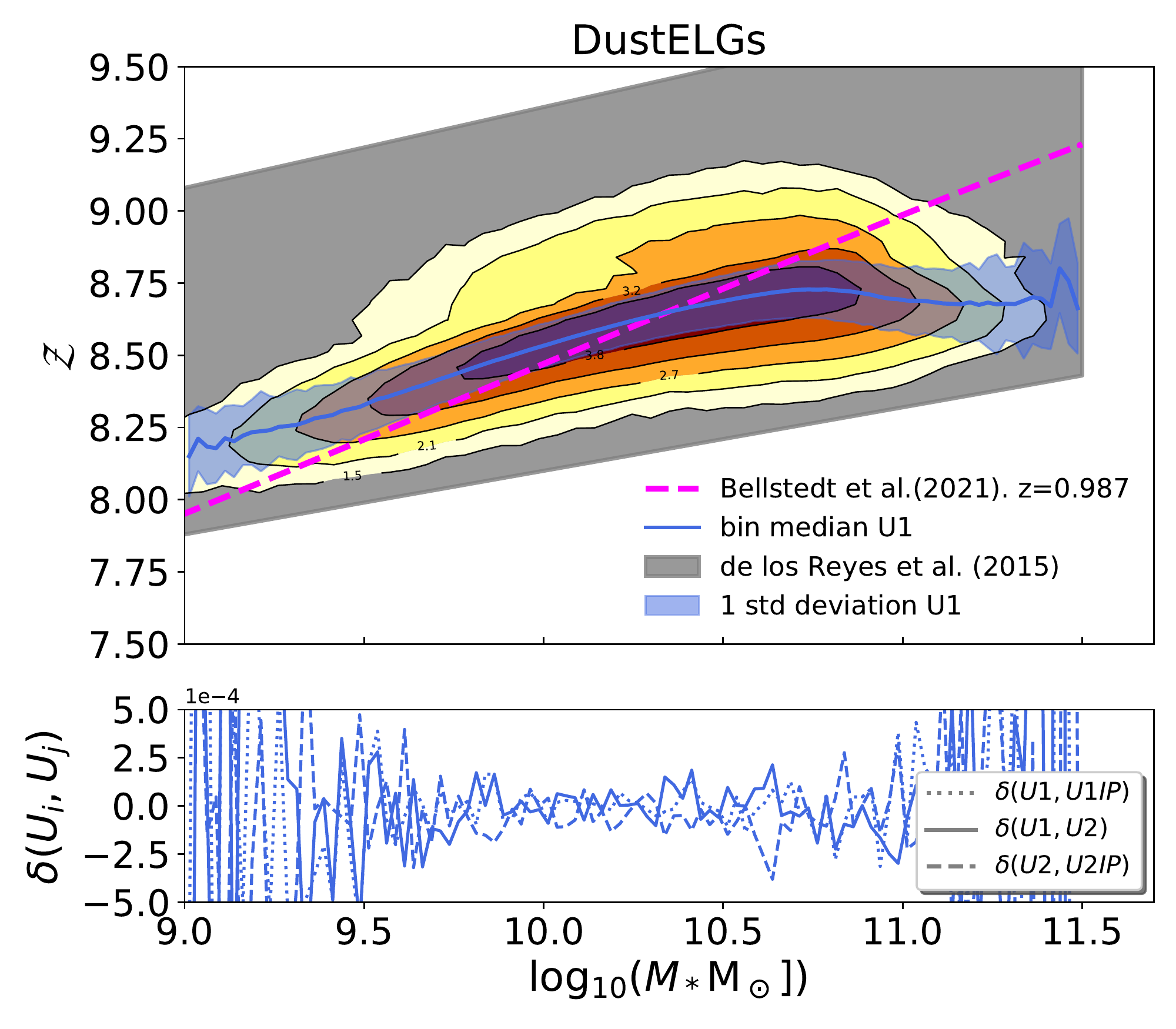}
   \caption{Cold gas metallicity vs. stellar mass for all ELGs for \RawMod\ (top) and \DustMod\ (bottom) at redshift $z\sim 1$. This figure is a reproduction of \Fig{fig:Zcold}, but this time for our model ELGs, but we also added the best-fitting relation as found by \citet{delosReyes2015} at $z\sim0.8$, shown as grey-shaded region.}
 \label{fig:ZcoldofELGs}
 \vspace{-0.3cm}
 \end{figure}

Here we reproduce \Fig{fig:Zcold} for the \RawMod\ and \DustMod\ catalogues, additionally adding the best-fit relation for \Halpha-emitting galaxies, as reported by \citet[][eq.~4]{delosReyes2015}. The results can be viewed in \Fig{fig:ZcoldofELGs}, which shows that the \sage-ELGs follow the observations sufficiently well.

\subsubsection{The disc size--mass relation} \label{app:RofELGs}
\begin{figure}
   \includegraphics[width=0.95\columnwidth]{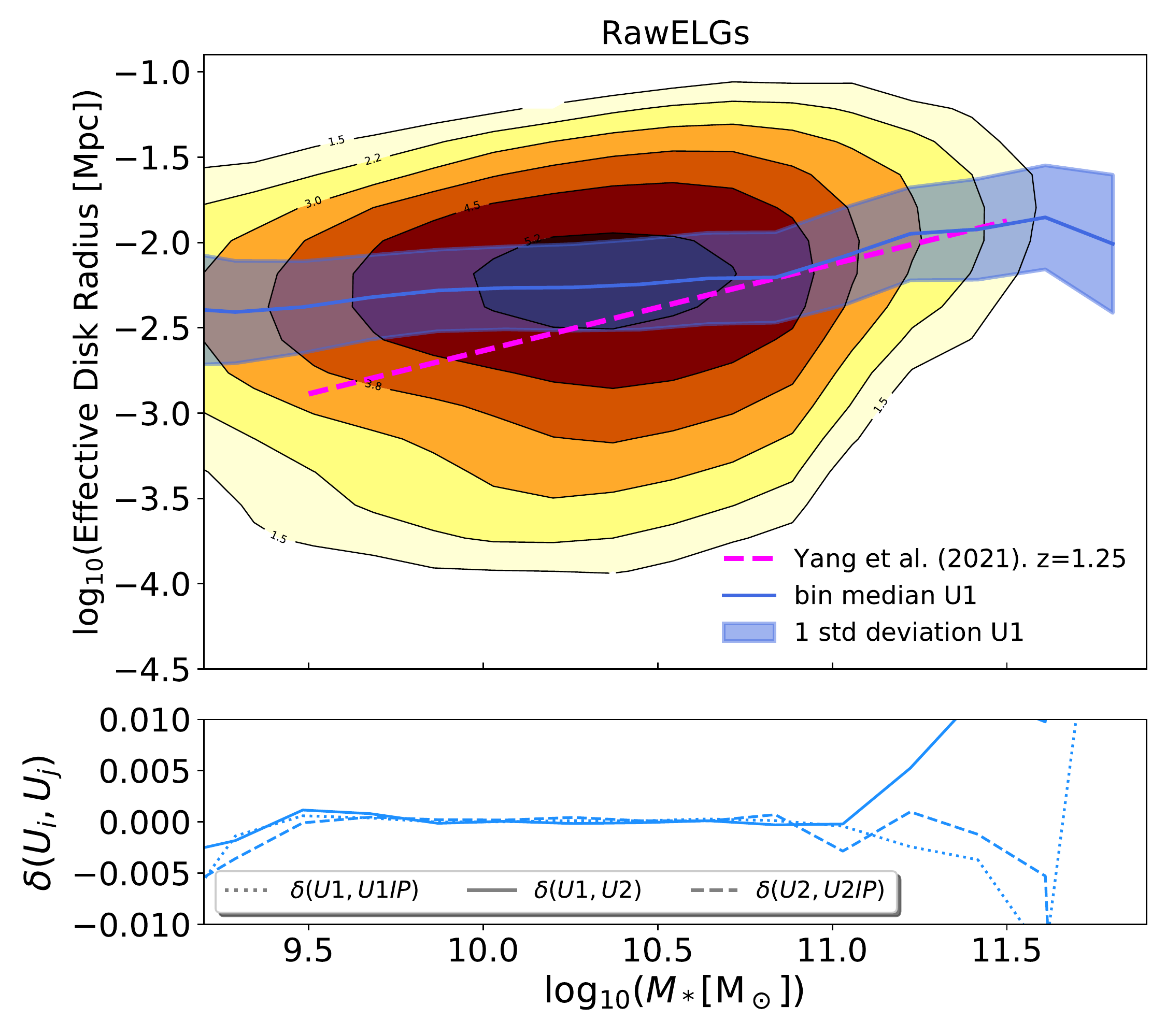}
   \includegraphics[width=0.95\columnwidth]{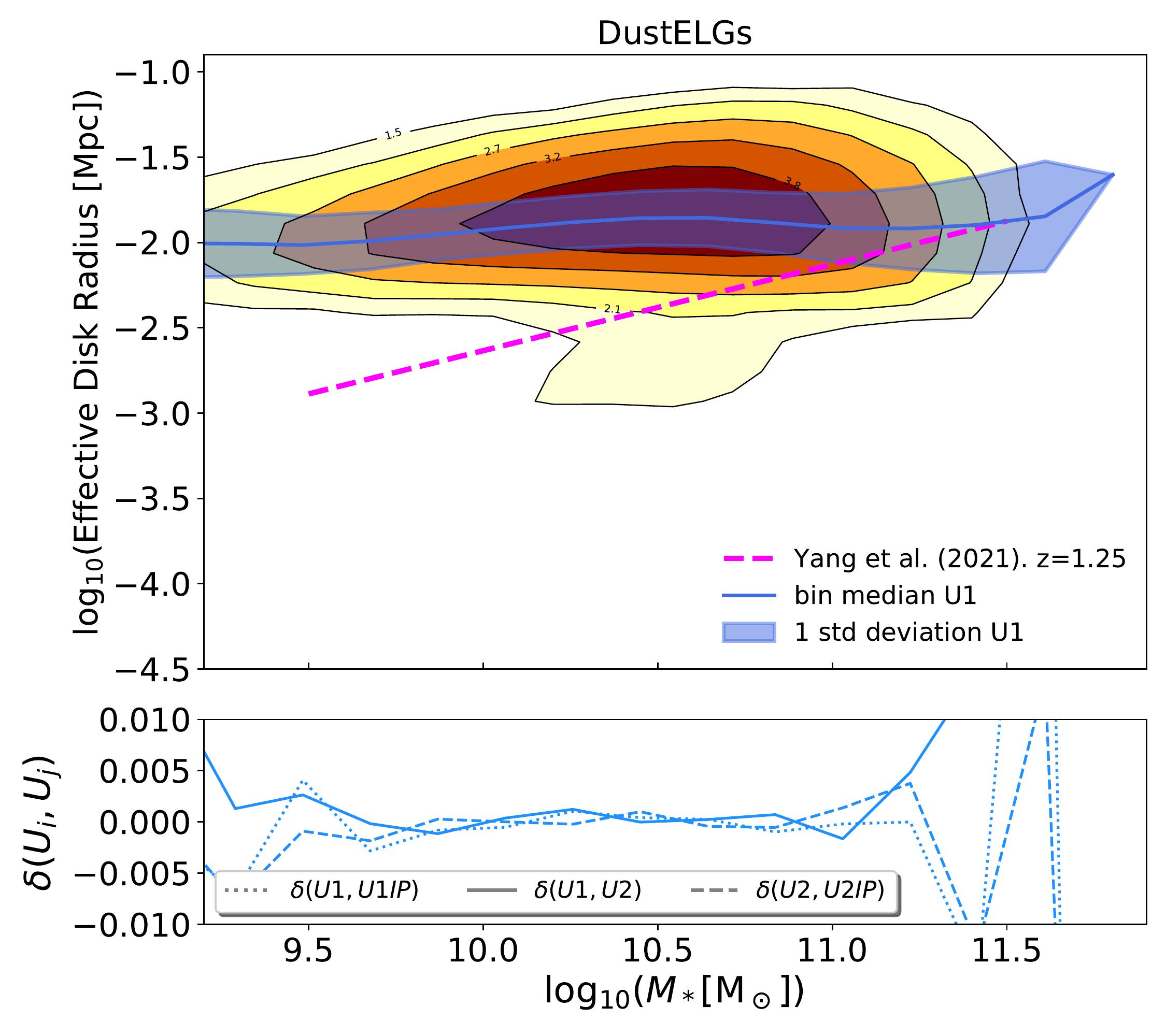}
   \caption{Effective disc radius as a function of stellar mass at redshift $z=1.25$ for all \RawMod\ (top) and \DustMod\ (bottom) galaxies. This figure is a reproduction of \Fig{fig:R_vs_SM}, but this time for our model ELGs.}
 \label{fig:R_vs_SMofELGs}
 \vspace{-0.3cm}
 \end{figure}

At last we turn to the effective disc size of our \RawMod\ and \DustMod\ galaxies, shown in \Fig{fig:R_vs_SM} for all \sage\ galaxies. The results can be viewed in \Fig{fig:R_vs_SMofELGs}, again in comparison to the general results of \citet{Yang2021}.

\section{Conversion of number densities} \label{app:conversion}
Here we show the steps necessary to go from volumetric number density
\begin{equation}
    n = \frac{dN}{dV}
\end{equation}

\noindent
to the angular and redshift density
\begin{equation}
    \eta = \frac{dN}{d\Omega\ dz} \ .
\end{equation}

\noindent
Taking into account 
\begin{equation}
    dV = d\Omega\ r^2 dr
\end{equation}

\noindent
where $d\Omega$ is the solid angle in stereoradians, we then get 
\begin{equation}
    \eta = n \cdot r^2 \frac{dr}{dz} \ .
\end{equation}

\noindent
Therefore, to go from number density $n=N/V$ of galaxies to number density of galaxies per square degree and redshift interval we find
\begin{equation} \label{eq:conversion}
    \eta = n \ r^2(z) \frac{dr}{dz} \left( \frac{\pi}{180^{\circ}}\right)^2 \ ,
\end{equation}

\noindent
where $r(z)$ is the comoving distance

\begin{equation}
    r(z) = \frac{c}{H_0} \int_{0}^{z} \frac{ds}{E(s)}
\end{equation}

\noindent
with

\begin{equation}
    E^2(z) = \frac{1}{(\Omega_{r,0} (1+z)^4 + \Omega_{m,0} (1+z)^3 + \Omega_{k,0} (1+z)^2 + \Omega_{\Lambda,0})} \ ,
\end{equation}

\noindent
where $\Omega_X$ are the usual density parameters of radiation ($X=r$), matter ($X=m$), curvature ($X=k$), and cosmological constant ($X=\Lambda$) at present time. We note that the derivative of $r(z)$ with respect to $z$ as needed in \Eq{eq:conversion} is simply

\begin{equation}
    \frac{dr}{dz} = \frac{c}{H_0} \frac{1}{E(z)}.
\end{equation}

\noindent
Note that in the main body of the paper $\eta$ is referred to as $dN/dz$, which is not fully consistent with the terminology used here, but compliant with how other workers in the field refer to this quantity. $N$ as used in the main part is `number of galaxies per unit area', whereas here it simply means `number of galaxies'. 

\bsp	
\label{lastpage}
\end{document}